\documentclass{emulateapj}

\usepackage{epsfig}
\usepackage{lscape}

\newcommand{\etal}{\mbox{et al.}}
\newcommand{\ergcms}{erg cm$^{-2}$ s$^{-1}$}
\newcommand{\ergs}{erg s$^{-1}$}
\newcommand{\phcms}{ph cm$^{-2}$ s$^{-1}$}
\newcommand{\degree}{$^\circ$}
\newcommand{\msun}{$M_{\odot}$}

\newcommand{\chandra}{{\it Chandra}}

\newcommand{\xmm}{{\it XMM-Newton}}

\newcommand{\fdeg}{\mbox{$.\!^{\circ}$}}

\newcommand{\sgrastar}{\mbox{Sgr A$^*$}}
\newcommand{\ebh}{\mbox{1E 1740.7--2942}}

\newcommand{\program}[1]{{\tt {#1}}}

\makeatletter

\newcommand\newtablebreak{\cr\ptable@@split}
\makeatother

\shortauthors{Muno \etal}
\shorttitle{X-ray Sources in the Nuclear Bulge}

\begin{document}

\title{A Chandra Catalog of X-ray Sources in the Central 150 pc of the Galaxy}
\author{M. P. Muno,\altaffilmark{1,2} 
F. E. Bauer,\altaffilmark{3,4} R. M. Bandyopadhyay,\altaffilmark{5,6}
and Q. D. Wang\altaffilmark{7,8}
}

\altaffiltext{1}{Department of Physics and Astronomy, University of California,
Los Angeles, CA 90095; mmuno@astro.ucla.edu}
\altaffiltext{2}{Hubble Fellow}
\altaffiltext{3}{Columbia Astrophysics Laboratory, Columbia University, Pupin
Laboratories, 550 W. 120th St., Rm 1418, NY, NY, 10027}
\altaffiltext{4}{Chandra Fellow}
\altaffiltext{5}{Dept. of Astrophysics, University of Oxford, Keble Road, 
Oxford OX1 3RH, U.K.}
\altaffiltext{6}{Dept. of Astronomy, University of Florida, 211 Bryant 
Space Science Center, Gainesville, FL, 32611  USA}
\altaffiltext{7}{Department of Astronomy, University of Massachusetts,
  Amherst, MA 01003; wqd@astro.umass.edu}
\altaffiltext{8}{Institute for Advanced Study, Einstein Drive, Princeton, NJ
08540}

\begin{abstract}
We present the catalog of X-ray sources detected in a shallow \chandra\ survey 
of the inner 2\degree$\times$0.8\degree\ of the Galaxy,
and in two deeper observations of the Radio Arches and Sgr B2. 
The catalog contains 1352 objects that are highly-absorbed 
($N_{\rm H}$$\ga$$4\times10^{22}$ cm$^{-2}$) and are therefore likely to 
lie near the Galactic center ($D$$\approx$8 kpc), and 549 less-absorbed 
sources that lie within $\la$6 kc of Earth.
Based on the inferred luminosities of the X-ray sources and the 
expected numbers of various classes of objects, we suggest that the 
sources with $L_{\rm X} \la 10^{33}$ \ergs\ that comprise $\approx$90\% 
of the catalog are cataclysmic 
variables, and that the $\approx$100 brighter objects are 
accreting neutron stars and black holes, young isolated pulsars, and 
Wolf-Rayet and O stars in colliding-wind binaries. We find that the spatial 
distribution of X-ray sources matches that of the old stellar population 
observed in the infrared, which supports our 
suggestion that most of the X-ray sources are old cataclysmic variables. 
However, we find that there is an apparent excess of $\approx$10 bright 
sources in the Radio Arches region. That region is already known to 
be the site of recent star formation, so we suggest that the bright sources 
in this region are young high-mass X-ray binaries, pulsars, or WR/O star 
binaries. We briefly discuss some astrophysical questions that this catalog 
can be used to address.
\end{abstract}

\keywords{catalogs --- Galaxy: center --- X-rays: general}

\section{Introduction}

The exquisite sensitivity of the {\it Chandra X-ray Observatory} has 
provided vast improvements in our understanding of faint, hard X-ray sources. 
\chandra\ observations of distant galaxies allow us to study the X-ray 
population at luminosities similar to those accessible in our own Galaxy 
with wide-field X-ray instruments like the {\it Rossi X-ray Timing 
Explorer} All-Sky Monitor and the {\it BeppoSAX} Wide-Field Camera 
($L_{\rm X} \sim 10^{36}$~\ergs), while observations of 
our own Galaxy are sensitive to sources a million times fainter than 
found with previous wide-field surveys. 
One of the most dramatic products of this improvement in sensitivity are
the \chandra\ observations of the Galactic center (Wang, Gotthelf, \& Lang
2002; Baganoff \etal\ 2003)\nocite{wgl02,bag03}.
Whereas previous imaging surveys identified dozens of X-ray sources against
a background of bright Galactic diffuse emission 
\citep{wat81,pav94,pt94,sbm01,sid99,sak02}, 
\chandra\ 
observations have revealed thousands of individual X-ray sources 
\citep[e.g.,][]{wgl02,mun03a} and discrete, filamentary features 
\citep[e.g.,][]{lwl03,yz05a}. 
This high concentration of X-ray sources is not surprising. The 
2\degree$\times$0\fdeg8 
\citep[300$\times$125 pc for a distance of 8 kpc;][]{mcn00} 
region centered on the Galactic 
center contains roughly 1\% of the Galactic mass (Lauhardt, Zylka, \& 
Mezger 2002)\nocite{lzm02}. 
Moreover, unlike the Galactic bulge, star formation has occurred continuously 
in the central region of the Galaxy, as is strikingly illustrated by 
the $\ga$60 ultra-compact HII regions in the
giant molecular cloud Sgr B2 \citep{dgg98}, 
and by three young, dense clusters of massive stars 
\citep[the Arches, the Quintuplet, and the Central Parsec;][]{kra95,fig99}.

\begin{figure*}[th]
\centerline{\epsfig{file=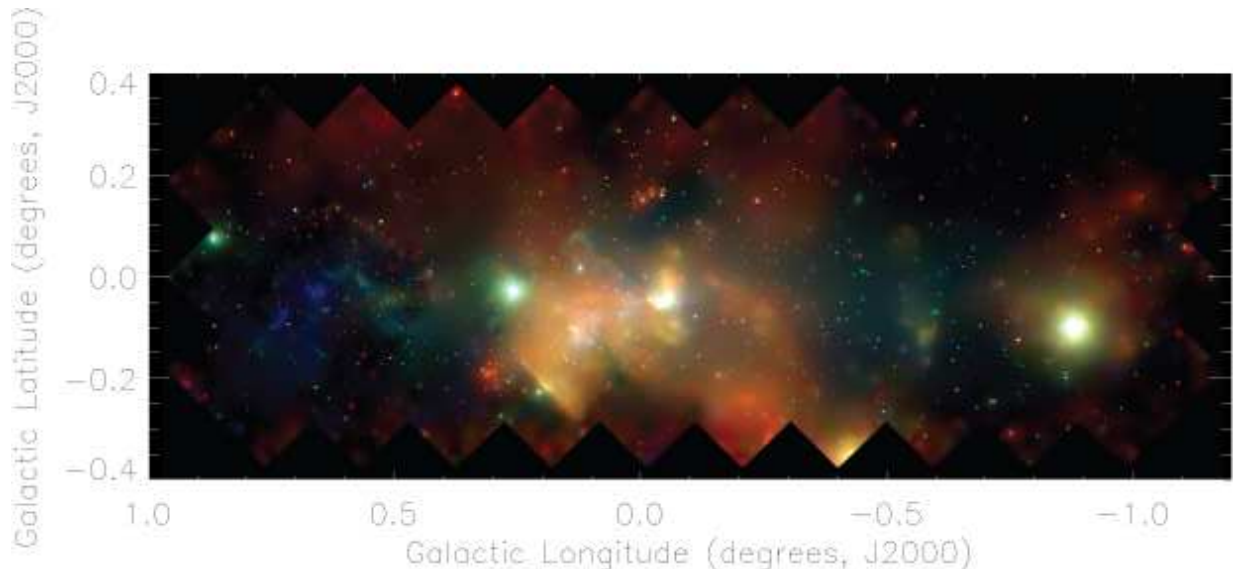,height=0.9\linewidth,angle=90}}
\caption{Mosaic image
of the 0\fdeg8$\times$2\degree\ field along the Galactic plane, 
centered on \sgrastar. The raw image has been adaptively smoothed 
with the CIAO tool {\tt csmooth} for display purposes. 
The prominent features are the Sgr A
complex at the center of the image, two bright X-ray binaries and 
their dust scattering halos at $l$=0.275\degree and $-$0.88\degree. 
Only a fraction of the brightest point sources are visible in this image.
Note that the smoothing algorithm introduces significant artifacts, 
especially at the edges of the deep observations centered on 
Sgr B2 and the Arches.
}
\label{fig:img}
\end{figure*}

A wealth of questions can be addressed with \chandra\ observations,
and with subsequent comparisons to multi-wavelength catalogs. For instance, a 
variety of studies of the synthesis of compact, accreting binaries 
have been designed to explain the large number of X-ray sources \chandra\ 
detects in the Galactic center, and the results constrain, for example, 
the amount of angular momentum dissipated in the common envelope phase
(e.g., Pfahl, Rappaport, \& Podsiadlowski 2002; Belczynski \& Taam 2004;
Liu \& Li 2005; Ruiter, Belczynski, \& Harrison 2005).
\nocite{prp02,bt04,ll05,rbh05}
\chandra\ also can detect outbursts from transient low-mass X-ray binaries 
(LMXBs) at much lower flux levels than are accessible with traditional 
wide-field 
X-ray surveys, which provides unique insight into the duty cycles 
and emission mechanisms of compact objects accreting at very low rates
\citep[$\dot{M} \la 10^{-11}$ \msun\ yr$^{-1}$; e.g.,][]{king00,wmw02,ww02,inz05,sak05}.
Combining \chandra\ and radio observations can be used to identify young 
stars with powerful winds, which helps to constrain the rate at which 
massive stars have formed recently in the Galactic center 
\citep{yz02,lyz04,mun05b}.

Several observational studies have discussed the population of X-ray 
sources in \chandra\ surveys of the Galactic center.
\citet{wgl02} presented images from shallow (12~ks) \chandra\ 
exposures of the 2\degree$\times$0\fdeg8 around \sgrastar, and gave
a general overview of the number of X-ray sources and the properties
of the diffuse X-ray emission. 
Takagi, Murakami, \& Koyama (2002)\nocite{tak02} 
studied the properties of X-ray sources associated with 
HII regions in Sgr B2. \citet{yz02} and \citet{lyz04} and studied the 
X-ray emission from massive stars in several clusters, including the Arches 
and Quintuplet. Finally, \citet{mun03a,mun04b,mun05a}, 
presented a comprehensive study of the population of X-ray sources with 
$L_{\rm X} = 10^{31} - 10^{33}$~\ergs\ that were discovered in a deep (625~ks) 
set of \chandra\ observations of the inner 25 pc around \sgrastar.
However, aside from the inner 25 pc, no catalog containing all of the X-ray 
sources found within the inner 150 pc of the Galactic center has been 
published.

In this paper, we rectify this by reporting the locations and basic 
properties of the X-ray sources detected in \chandra\ observations of 
the inner 2\degree$\times$0\fdeg8 around the Galactic center. 
In \S\ref{sec:id}, we present the locations of these X-ray sources 
\citep[excluding the inner 8\arcmin\ covered by the catalog in][]{mun03a}
in order to facilitate searches for 
multi-wavelength counterparts. In \S\ref{sec:phot}, 
we present the fluxes and basic spectral properties in order to constrain
the origin of the X-ray emission, and to serve as baseline measurements 
for future searches for transient sources. In \S\ref{sec:dist}, we 
examine the spatial distribution of the X-ray sources to determine how 
they are related to the stellar population that is observed in 
the infrared. In \S\ref{sec:ns}, we study the luminosity distribution of
the X-ray sources, and report variations in the relative numbers of 
bright sources that could be related to recent star formation.

\section{Observations}

The central 300$\times$125 pc of the Galaxy
has been observed on several occasions with the imaging array of the 
\chandra\ Advanced CCD Imaging Spectrometer \citep[ACIS-I][]{wei02}. 
The entire region was surveyed with overlapping $\approx$12~ks observations
\citep{wgl02}, and one additional short observation was obtained 
centered on the LMXB \ebh. Deeper observations were taken 
of the HII regions in the giant molecular cloud Sgr B2 
\citep[100~ks;][]{tak02}, and the non-thermal radio features referred to 
as the Arches \citep[50~ks of public data as of 2005 June;][]{lyz04}.
We list each of these observations in Table~\ref{tab:obs}, and present 
a mosaic image of the survey in Figure~\ref{fig:img}.

\begin{deluxetable*}{lclcccc}[thb]
\tabletypesize{\scriptsize}
\tablecolumns{7}
\tablewidth{0pc}
\tablecaption{Observations of the Central 2\degree$\times$0.8\degree of the Galaxy\label{tab:obs}}
\tablehead{
\colhead{} & \colhead{} & \colhead{} & \colhead{} & 
\multicolumn{2}{c}{Aim Point} & \colhead{} \\
\colhead{Start Time} & \colhead{Sequence} & \colhead{Target} & 
\colhead{Exposure} & \colhead{RA} & \colhead{DEC} & \colhead{Roll} \\
\colhead{(UT)} & \colhead{} & \colhead{} & \colhead{(ks)} 
& \multicolumn{2}{c}{(degrees J2000)} & \colhead{(degrees)}
} 
\startdata
2000 Aug 30 16:59:32 & 658 & 1E 1740.7$-$2942 & 9.2 & 265.97583 & $-$29.75008 & 270.8 \\
2000 Mar 29 09:44:36 & 944 & SGR B2 & 97.5 & 266.78034 & $-$28.44169 & 87.8 \\
2000 Jul 07 19:05:19 & 945 & GC ARC & 48.8 & 266.58192 & $-$28.87196 & 284.4 \\
2001 Jul 19 10:01:48 & 2267 & GCS 20 & 8.7 & 266.17150 & $-$29.27337 & 283.8 \\
2001 Jul 20 04:37:11 & 2268 & GCS 21 & 10.8 & 265.98136 & $-$29.17141 & 283.8 \\ [5pt]
2001 Jul 16 02:15:50 & 2269 & GCS 1 & 10.5 & 267.05495 & $-$28.37576 & 283.8 \\
2001 Jul 20 08:00:49 & 2270 & GCS 22 & 10.6 & 266.24512 & $-$29.54138 & 283.8 \\
2001 Jul 16 05:35:55 & 2271 & GCS 2 & 10.4 & 266.86502 & $-$28.27455 & 283.8 \\
2001 Jul 20 11:12:40 & 2272 & GCS 23 & 11.6 & 266.05423 & $-$29.43957 & 283.8 \\
2001 Jul 18 00:48:28 & 2273 & GCS 10 & 11.2 & 266.70988 & $-$28.87565 & 283.8 \\ [5pt]
2001 Jul 16 08:44:25 & 2274 & GCS 3 & 10.4 & 266.67662 & $-$28.17301 & 283.8 \\
2001 Jul 20 14:41:10 & 2275 & GCS 24 & 11.6 & 265.86371 & $-$29.33729 & 283.8 \\
2001 Jul 18 04:16:58 & 2276 & GCS 11 & 11.6 & 266.51970 & $-$28.77438 & 283.8 \\
2001 Jul 16 11:52:55 & 2277 & GCS 4 & 10.4 & 266.94061 & $-$28.54231 & 283.8 \\
2001 Jul 20 18:09:40 & 2278 & GCS 25 & 11.6 & 266.12769 & $-$29.70775 & 283.8 \\ [5pt]
2001 Jul 18 07:45:28 & 2279 & GCS 12 & 11.6 & 266.33020 & $-$28.67281 & 283.8 \\
2001 Jul 16 15:01:25 & 2280 & GCS 5 & 10.4 & 266.75037 & $-$28.44124 & 283.8 \\
2001 Jul 20 21:38:10 & 2281 & GCS 26 & 11.6 & 265.93652 & $-$29.60557 & 283.8 \\
2001 Jul 18 11:13:58 & 2282 & GCS 13 & 10.6 & 266.59425 & $-$29.04216 & 283.8 \\
2001 Jul 21 01:06:39 & 2283 & GCS 27 & 11.6 & 265.74584 & $-$29.50315 & 283.8 \\ [5pt]
2001 Jul 18 14:25:48 & 2284 & GCS 14 & 10.6 & 266.40487 & $-$28.94088 & 283.8 \\
2001 Jul 16 18:09:55 & 2285 & GCS 6 & 10.4 & 266.56112 & $-$28.34029 & 283.4 \\
2001 Jul 21 04:35:09 & 2286 & GCS 28 & 11.6 & 266.00997 & $-$29.87372 & 283.8 \\
2001 Jul 18 17:37:38 & 2287 & GCS 15 & 10.6 & 266.21439 & $-$28.83925 & 283.8 \\
2001 Jul 17 14:11:51 & 2288 & GCS 7 & 11.1 & 266.82518 & $-$28.70891 & 283.8 \\ [5pt]
2001 Jul 21 08:03:39 & 2289 & GCS 29 & 11.6 & 265.81855 & $-$29.77165 & 283.8 \\
2001 Jul 21 11:32:10 & 2290 & GCS 30 & 11.6 & 265.62772 & $-$29.66900 & 283.8 \\
2001 Jul 18 20:49:28 & 2291 & GCS 16 & 10.6 & 266.47839 & $-$29.20880 & 283.8 \\
2001 Jul 17 17:51:28 & 2292 & GCS 8 & 11.6 & 266.63516 & $-$28.60795 & 283.8 \\
2001 Jul 19 00:01:18 & 2293 & GCS 17 & 11.1 & 266.28794 & $-$29.10740 & 283.8 \\ [5pt]
2001 Jul 17 21:19:58 & 2294 & GCS 9 & 11.6 & 266.44581 & $-$28.50671 & 283.8 \\
2001 Jul 19 03:21:28 & 2295 & GCS 18 & 11.1 & 266.09836 & $-$29.00518 & 283.8 \\
2001 Jul 19 06:41:38 & 2296 & GCS 19 & 11.1 & 266.36205 & $-$29.37522 & 283.8
\enddata
\end{deluxetable*}

The ACIS-I is a set of four 1024-by-1024 pixel CCDs, covering
a field of view of 17\arcmin\ by 17\arcmin. When placed on-axis at the focal
plane of the grazing-incidence X-ray mirrors, the imaging resolution 
is determined primarily by the pixel size of the CCDs, 0\farcs492. 
The CCDs also 
measure the energies of incident photons within a 
calibrated energy band of 0.5--8~keV, with a resolution of 50--300 eV 
(depending on photon energy and distance from the read-out node). 
The CCD frames are read out every 3.2~s,
which provides the nominal time resolution of the data.

We reduced the observations using standard tools from the 
CIAO package, version 2.3.01. We started with the level-1 event lists provided
by the \chandra\ X-ray Center (CXC), and removed the pixel randomization 
applied by the default processing software. We then modified the pulse 
heights of each event to partially correct for the position-dependent
charge-transfer inefficiency caused by radiation damage early in the 
mission, using software provided by \citet{tow02b}.  We excluded most events
flagged as possible background, but left in possible cosmic ray
afterglows because in the version of the processing software that we used they 
were difficult to distinguish from genuine X-rays 
from the strong diffuse emission and numerous point 
sources in the field. We applied the standard ASCA grade filters to the 
events, as well as the good-time filters supplied by the CXC. Finally,
we searched each observation for time intervals when the detector 
background flared to $\ge 3\sigma$ above the mean level, and removed
such intervals when they occurred (in ObsIDs 2267, 2269, 2273, 2288, and 944). 

\subsection{Source Detection and Initial Localization\label{sec:id}}

We searched for X-ray sources separately in sets of 9 images for each 
observation using the wavelet routine \program{wavdetect} 
\citep{free02}.
We generated images in three energy bands: the full 0.5--8.0 keV band,
the 0.5--2.0 keV band to increase our sensitivity to foreground sources,
and 4--8 keV to increase our sensitivity to highly absorbed sources. 
For the purposes of 
source detection only, we removed events that had been flagged as 
possible cosmic ray afterglows. We employed the default 
``Mexican Hat'' wavelet, and used a sensitivity threshold of $10^{-7}$
that corresponds to chance of detecting a spurious source
per pixel if the local background is spatially uniform.
We searched each energy band 
using a succession of three images centered on the aim point
of each exposure: 1024$\times$1024 images at the full \chandra\
resolution of 0\farcs5, 1024$\times$1024 images binned by a factor of 2 
to a resolution of 1\arcsec, and
images of variable size that covered the entire ACIS-I exposure for
each observation with a resolution of 2\arcsec. We used wavelet scales 
that increased by a factor of $\sqrt{2}$: 1--4 for the 0\farcs5 image, 
1--8 for the 1\arcsec\ image, and 1--16 for the 2\arcsec\ image.
This succession of three images and spatial scales were chosen because the 
point-spread function (PSF) broadens as a function of offset from the 
aim point.
The resulting source list from the 33 observations contained 1901
unique sources, 225 of which were only detected in the soft band, and 382
of which were only detected in the hard band. Based on the sensitivity
threshold for {\tt wavdetect} ($10^{-7}$), we expect 2 spurious sources
per field, or $\sim$70 in the entire survey.

We attempted to refine the astrometry for each observation by matching
foreground X-ray sources detected in the soft band, many of which are
likely to be K and M dwarf stars \citep{mun03a}, to infrared sources in
the Two-Micron All-Sky Survey (2MASS) catalog. For the two observations
longer than 50~ks (Sgr B2 and the Arches), we found that there were 
$\approx$20  matches between the soft X-ray and 2MASS catalog within 
5\arcmin\ of the aim point. By randomly shifting the relative positions of 
the two catalogs, we determined that, with 90\% confidence, fewer than 
25\% of these matches should be random. Based on these matches, we could 
derive the absolute astrometry of the \chandra\ pointing to within 
0\farcs1. 

Unfortunately, the shorter observations were less sensitive, and contained
far fewer X-ray sources. Whereas the $\ga 50$~ks exposures contained
$\approx$200 X-ray sources, the shorter exposures generally contained only 
a couple dozen. In general, $\le$2 X-ray sources could be identified 
with 2MASS counterparts, which was insufficient to improve the pipeline 
astrometry. 
Therefore, starting with the fields adjacent to the deep exposures of
\sgrastar\ \citep{mun03a}, the Arches, and Sgr B2
and moving outward, we derived the astrometry by matching the 
\program{wavdetect} positions of the X-ray sources matched those of 
adjacent fields. Unfortunately, because most of the matches made in this
way relied on X-ray sources detected $\ga 6$\arcmin\ off-axis, the 
statistical uncertainty on the \program{wavdetect} positions of each 
X-ray sources was significant, and the corrected astrometry again was 
not significantly more accurate than the default pipeline values. 
Therefore, we expect that 
the astrometry will on average be accurate to 0\farcs7, but that $\approx$10\%
of the pointings will have systematic errors of up to 1\arcsec. Unfortunately, 
there is no way of knowing {\it a priori} which observations have
the larger astrometric errors, so our astrometric accuracy is limited
to $\approx$1\arcsec\ in the shallower exposures. 

In Table~\ref{tab:cat}, we list the refined positions and  
positional uncertainties (90\% confidence), which include
statistical and systematic terms. For the statistical positional error,
we use the count-rate-dependent estimates in \citet[][]{alex03}. For 
the majority of sources with $<$200 net counts, we assume the 
uncertainty $\Delta$=0\farcs6 for offsets $\theta$$\le$5\arcmin\ and
\begin{equation}
\Delta = 0.6 + \left({{\theta - 5'}\over{6.25'}}\right)~{\rm arcsec}
\end{equation}
for $\theta$$>$5'. For 19 bright sources with $\ge$200 net counts, 
we assume $\Delta$=0\farcs3 for offsets $\theta$$\le$5\arcmin\ and
\begin{equation}
\Delta = 0.3 + \left({{\theta - 5'}\over{25'}}\right)~{\rm arcsec}
\end{equation}
for $\theta$$>$5'.
We combine the statistical errors in quadrature with 0\farcs7 systematic 
uncertainties for the shallow observations 
(all those except ObsIDs 944 and 945, for which we assume no systematic
uncertainty).
In Table~\ref{tab:cat}, we also list the observation in which a source 
was detected in, the offset from the aim point of that observation, 
and the total live time each source was observed with. We omitted  
two bright, previously-known LMXBs from the table
(\ebh\ and 2E 1743.1--2842), because they were badly saturated in our 
survey images, and their positions and properties are reported elsewhere 
\citep[e.g.,][]{mar00,por03}.

\subsection{Photometry\label{sec:phot}}

We computed photometry for each source in the 0.5--8.0~keV band using 
the \program{acis\_extract} routine from the Tools for 
X-ray Analysis (TARA).\footnote{{\tt www.astro.psu.edu/xray/docs/TARA/}}
We extracted event lists for each source for each observation, using an
extraction region designed to enclose a large fraction of the PSF.
We used a PSF at a fiducial energy of 1.5~keV
for sources detected only in the soft band (flagged with a 'f' to
indicate a possible foreground source in Tab.~\ref{tab:cat}), while we 
used a larger extraction area corresponding to a PSF for 4.5 keV
photons when sources were detected in the full or hard bands. 

In most cases, we chose
polygonal regions that matched the contours of 90\% encircled energy
from the PSF. However, if the 90\% contours of the PSFs of
two nearby sources overlapped, we generally used a region that 
corresponded to a smaller fraction of the PSF. 
We have flagged these sources as confused ('c') in 
Table~\ref{tab:cat}.
The smallest extraction region that we used matched the 70\% encircled 
energy contour. However, because the 
PSF grows significantly as a function of off-axis angle, in many 
cases a source that appeared isolated when it was located on-axis in one 
observation was indistinguishable from its neighbors in images from 
adjacent, overlapping pointings. We found that no reasonable fraction of the
PSF would isolate counts from sources more than 7\arcmin\ off-axis whose
90\% encircled-energy radii overlapped those of their neighbors, so we
did not extract photometry from observations in which confused sources 
were $>$7\arcmin\ off-axis.
Fortunately, these sources always lay near the aim point of another 
observation (or else they could not have been detected in the first place), 
so we were still able to obtain photometry for these sources. 

For each source and each observation, a background event list was 
extracted from a circular region centered on the point source, excluding 
from the event list counts in circles circumscribing 
the $\approx$92\% contour of the PSF around any point sources. We found that
this value struck a good balance between excluding counts from point sources
and leaving enough counts in the image to determine the background. 
The sizes of the 
background regions were chosen such that they contained approximately 100 
events for each observation. Less than 1\% of the counts in the background
regions were from point sources. We also computed the effective area function 
(ARF) at the position of each source for each observation using the CIAO 
tool \program{mkarf}.

\begin{figure}
\centerline{\epsfig{file=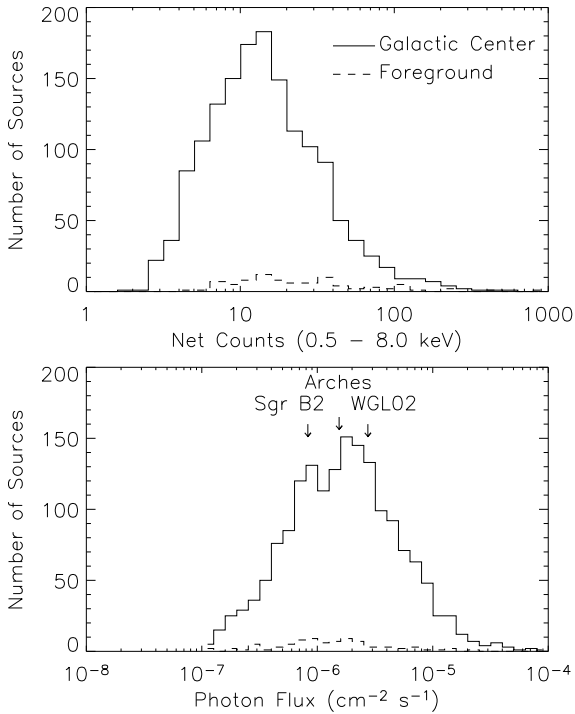,width=0.95\linewidth}}
\caption{{\it Top panel:} The distribution in net counts from individual
sources. No correction is applied to account for the exposure across the 
survey, which varies by a factor of 5. {\it Bottom panel:} The distribution
of fluxes from individual sources, derived by dividing the net count 
rates by the effective area and exposure in four energy bands (0.5--2.0~keV, 
2.0--3.3~keV, 3.3--4.7~keV, and 4.7--8.0~keV), and summing the result.
There are two peaks, because the deeper observations were more sensitive
to faint sources. In both panels, the solid line is used for sources
located near or beyond the Galactic center ($HR$$>$$-0.175$), and the 
dashed line for foreground sources ($HR$$<$$-0.175$). The arrows denote
the median sensitvity for the shallow survey (labeled WGL02) and for the 
deep Arches and Sgr B2 fields.
}
\label{fig:count}
\end{figure}

The source and background event lists were used to compute photometry for 
each source in five energy bands: 0.5--8.0 keV (the full band), 
0.5--2.0~keV, 2.0--3.3~keV, 3.3--4.7~keV, and 4.7--8.0~keV. 
These bands are identical to those 
used by \citet{mun03a} for the catalog of X-ray sources within 25 pc
of \sgrastar. The energy bands were chosen so
that they sampled regions of the ARF with roughly constant areas, and
so that the three high energy bands each contained about one-third of 
the net counts
from most sources. The net counts in each band were computed by subtracting 
the estimated background from total counts.
The 90\% uncertainty in the net counts in each of the five bands
were computed through a Bayesian analysis, with the simplifying assumption 
that the uncertainty on the background was negligible \citep{kbn91}. 
When the 90\% confidence interval on the net counts was consistent with 0, 
we considered the upper bound on the 90\% confidence interval to be 
the upper limit. 

\begin{figure*}[thb]
\centerline{\epsfig{file=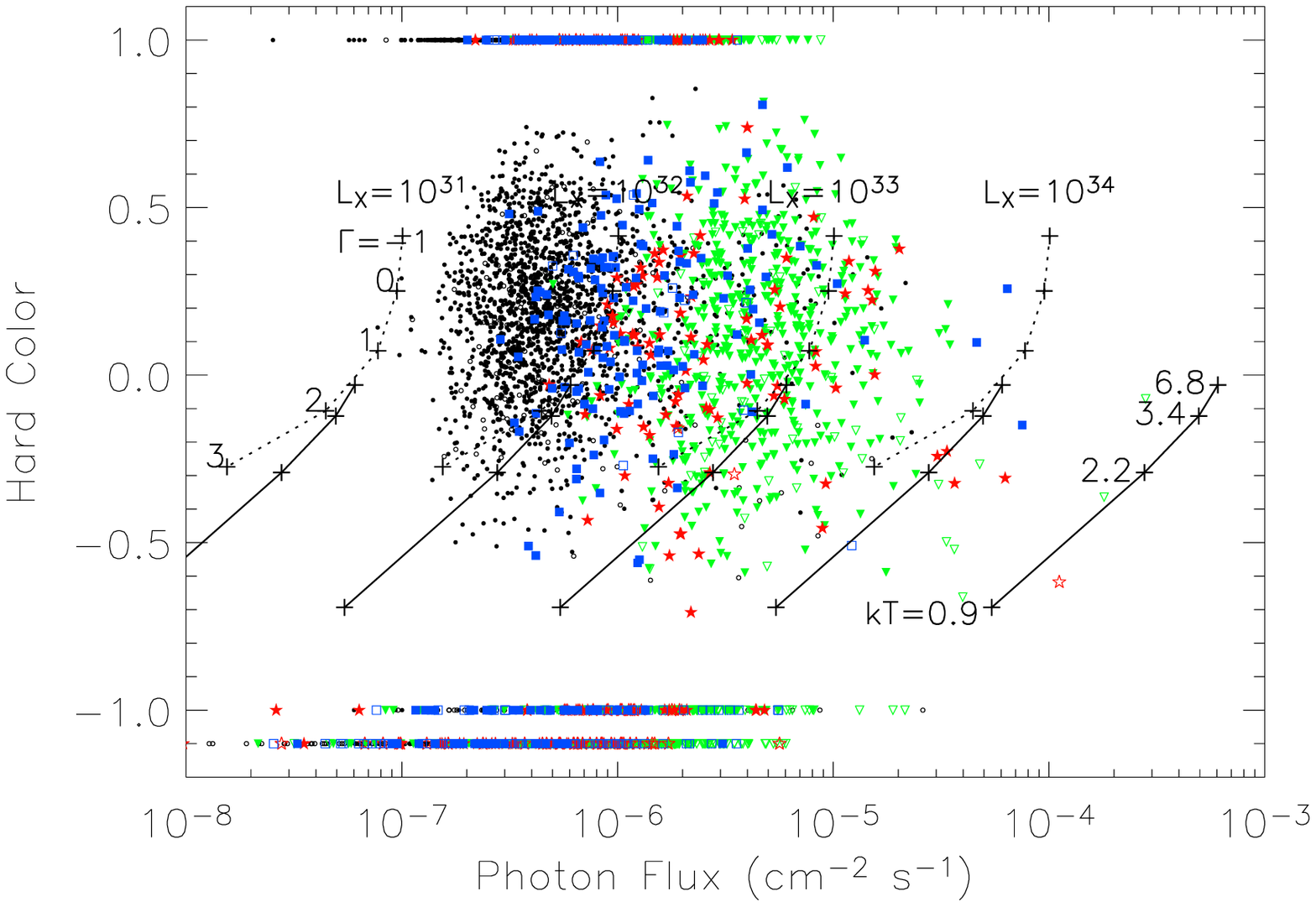,width=0.95\linewidth}}
\caption{The hard color plotted against the photon flux from
each source. The symbol shapes indicate which observations the sources
were identified in: blue squares for the Sgr B2 field, red stars for
the Arches region, and green triangles for the extended shallow survey.
For comparison, we also include the data from the central 25 pc
\citep{mun03a} using black circles. Open symbols indicate foreground
sources, and filled circles those at or beyond the Galactic center 
(see text).
Sources detected in only in the 3.3--4.7 keV band are assigned hard
colors of $-$1; those only detected in the 4.7--8.0 keV band are assigned
$HR2$=$+$1, and those detected in neither band are assigned $HR2$=$-$1.1.
The uncertainties on the hard colors are significant. Sources with a 
probability of $<$50\% of being detected have unreliable hard colors;
those with a 50--90\% chance of detection have $\sigma_{HR2} \approx 0.6$;
those with a 90--99\% chance of detection have $\sigma_{HR2} \approx 0.4$;
and those with a $>$99\% chance of detection have $\sigma_{HR2} \la 0.3$.
Finally, we have plotted the colors expected for sources of 
varying luminosities at a distance of 8 kpc, and absorbed by 
$6\times10^{22}$ cm$^{-2}$ of interstellar gas and dust. The dotted
lines are for power-law spectra, and the solid lines for thermal plasma
spectra. The sources with $HR2$$>$0.5 either have large uncertainties, large 
absorption columns, or both.
For a fiducial conversion factor between photon flux and
0.5--8.0 keV luminosity, we assume a $\Gamma$=1.5 power law or a
$kT$$=$7 keV plasma, 
and find that $10^{34}$ \ergs\ equals $6\times10^{-5}$ \phcms.
}
\label{fig:hit}
\end{figure*}

A histogram of the number of sources as a function of 
net counts in the full band is displayed in Figure~\ref{fig:count}.
Most sources were detected with 10--20 counts in two overlapping 12~ks 
observations. Therefore, most sources were only observed with 
$\approx$5 counts in each of the smaller energy bands. 
The net counts from 303 sources are consistent with zero at the 90\% level.
About 70\% of these sources are flagged in Table~\ref{tab:cat} because their
photometry could be unreliable for various reasons: some are detected only 
in the soft band where the background is lower, some are variable so that 
the mean flux is not meaningful, and some are confused with nearby sources. 
The remaining $\approx$90 sources are probably spurious, as we expected 
$\approx$70 spurious sources based on our detection threshold. We suspect we
detected $\approx$20 more spurious than we expected because the background 
in our observations is $\approx$10 times larger than in the 
observations taken at high Galactic latitude with which {\tt wavdetect} was 
calibrated. We keep these potentially spurious sources in the catalog for 
completeness. The net counts and 90\% uncertainties (or upper limits) are 
listed in Table~\ref{tab:cat}. These values are
used to compute the probability of detecting each source given its location 
and exposure, which is also listed in the table. The Monte Carlo simulations 
used to estimate this probability are described in Appendix A.

We computed approximate photon fluxes (in units of \phcms) for each source
by dividing the net counts in each sub-band by the total live time
(units of s) and the mean value of the ARF in that energy range 
(units of cm$^2$; note that this value incorporates variations in exposure due 
to chip gaps and dead columns). The photon 
fluxes in the 0.5--8.0~keV energy band used throughout the paper are the 
sums of those in the sub-bands, using negative values when they occur 
(not the upper limits). They are listed in Table~\ref{tab:cat}.
We have found that the approximate photon 
fluxes that we computed differed from those derived from spectral fits
by little more than the Poisson uncertainty in the count rate, because the 
energy bands sampled the ARF for the ACIS-I detector well \citep{mun03a}.

A histogram of the number of sources as a function 
of the 0.5--8.0~keV photon flux is presented in the {\it bottom panel} of 
Figure~\ref{fig:count}. Galactic center sources are indicated with the 
{\it solid line}, and foreground sources with the {\it dashed line}.
Sources are detected with average
photon fluxes as low as $2\times10^{-8}$~\phcms. The largest number of 
Galactic center sources are detected near $2\times10^{-6}$~\phcms\
(2.0--8.0~keV), and the largest number of foreground sources are found near 
$1\times10^{-6}$~\phcms\ (0.5--2.0~keV). 


We used the counts in each energy band to compute two hardness ratios,
which we used to characterize the absorption column toward each 
source and the steepness of the high-energy portion of each spectrum. 
The ratios are defined 
as the fractional difference between the count rates in two 
energy bands, $(h-s)/(h+s)$, where 
$h$ and $s$ are the numbers of counts in the higher and lower energy bands, 
respectively. The resulting ratio is bounded by $-1$ and $+1$. 
The soft color is defined by the fractional difference between counts 
with energies between 2.0--3.3~keV 
and 0.5--2.0~keV, and the hard 
color using counts between 4.7--8.0~keV and 3.3--4.7~keV. The hardness 
ratios are listed in Table~\ref{tab:cat}, with uncertainties calculated 
according to Equation~1.31 in Lyons (1991; page 26). 

We use the soft color ($HR0$) to identify foreground and 
highly-absorbed sources. By comparing the spectral fits and hard colors
in \citep{mun03a,mun04b}, we find that sources with $HR0 > -0.175$ or
that are not detected below 3.3~keV 
have absorption columns $N_{\rm H} > 4\times10^{22}$ cm$^{-2}$, and
are therefore likely to lie at or beyond the Galactic center. We 
refer to these as the ``Galactic center sources,'' of which there
are 1350. Sources
with $HR0< -0.5$ have
$N_{\rm H} \la 10^{21}$ cm$^{-2}$ and lie within
2~kpc of Earth. Sources with intermediate soft colors lie between 
2--6~kpc from the Earth. The 549 sources with $HR0 < -0.175$ that are 
likely to be closer than 6 kpc are considered foreground sources.

In Figure~\ref{fig:hit}, we plot the hard color versus the flux from
each source. Foreground sources are indicated with open circles, and 
sources at or beyond the Galactic center with filled circles.
There are 785 Galactic center sources and 39 foreground sources with 
measured hard colors. 
We have calculated the hardness ratios and photon fluxes that we would expect
to get from these energy bands for a variety of spectra and 0.5--8.0 keV
luminosities using \program{PIMMS} and \program{XSPEC}. In 
Figure~\ref{fig:hit}, we
plot the colors and fluxes expected for power-law spectra with the 
dotted lines, and for a optically-thin thermal plasma with the solid lines.
We have assumed a distance of 8 kpc and $6\times10^{22}$ cm$^{-2}$ of
absorption from interstellar gas and dust.
The median hard color for the Galactic center sources is $-$0.05. This 
corresponds to a $\Gamma$=1.5 power law or a $kT$=1.7 keV plasma 
spectrum. 

The Galactic center sources in this survey are significantly softer 
than those from the deeper (limiting luminosity of 
$L_{\rm X} = 2\times10^{31}$~\ergs) catalog from the central 20 pc
of the Galaxy. The latter catalog had a median hard color of 
$-0.22$ \citep{mun03a}, corresponding to a $\Gamma$=0 power law 
spectrum. This suggests that the more luminous X-rays sources are 
systematically softer.

For a fiducial $\Gamma = 1.5$ spectrum absorbed by 
$N_{\rm H}$$=$$6\times10^{22}$ cm$^{-2}$), 
the photon fluxes can be converted to energy fluxes according to
1~\phcms~$= 8\times10^{-9}$~\ergcms\ (0.5--8.0~keV). The de-absorbed 0.5--8.0
keV flux is approximately 3 times larger, so that for a distance
$D$=8 kpc, $10^{34}$ \ergs\ equals $6\times10^{-5}$ \phcms.
For sources detected below 2.0~keV, we
find that 1~\phcms~$= 2\times10^{-9}$ \ergcms\ between 0.5--2.0~keV. 
The absorption for these sources is relatively small 
($< 10^{22}$ cm$^{-2}$), and therefore so is the correction to derive an 
intrinsic flux.

\section{Results}

\subsection{Spatial Distribution\label{sec:dist}}

We examined the spatial distribution of the X-ray sources in order
to determine how it compares to that of the ordinary stellar population 
observed in the infrared. We were most interested in sources near the 
Galactic center, so we only considered those sources with soft colors
$HR0 > -0.175$. We also took care to select only sources that were 
bright enough to be detected over a large fraction of the survey. 
To do this, we derived maps of our sensitivity as described in 
Appendix A, and we examined only those sources that (1) were brighter than 
a well-defined flux limit, and (2) that were located at a 
position where the sensitivity was lower than that flux limit. 
Our flux limit was designed so that sources brighter than the limit 
have at least a 50\% chance of being detected over a significant fraction 
of our survey (see Appendix A for a description of the Monte Carlo 
simulations that we used to calculate the detection probability). 
We found that a flux limit of $3\times10^{-6}$ \phcms\ 
(equivalent to $5\times10^{32}$~\ergs\ [0.5--8.0 keV] taking $D$=8 kpc, 
$N_{\rm H} = 6\times10^{22}$cm$^{-2}$, and a $\Gamma$=1.5 power law) 
provided the largest number of sources, 321 out of the 1899 sources in 
Table~\ref{tab:cat}.

\begin{figure}
\centerline{\epsfig{file=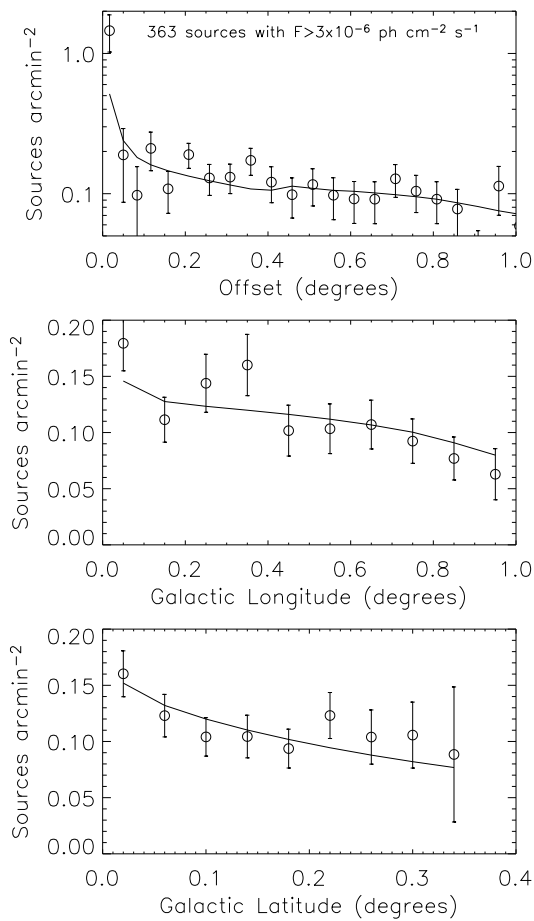,width=0.85\linewidth}}
\caption{
The distributions of point sources as a function of angular offset from 
\sgrastar\ ({\it top panel}), and the absolute values of Galactic 
longitude ({\it middle panel}) and latitude ({\it bottom panel}). We only 
considered sources that were brighter than $F_{\rm X} = 3\times10^{-6}$
\phcms, that had
a 50\% chance of being detected, and that lay in regions where the 
50\% detection threshold was lower than the above flux limit. The contribution
of each source to the distribution was weighted by the inverse of the 
probability of detecting it.
The surface density of 
stellar mass is plotted with the solid line, which has been normalized 
through a chi-squared minimization to match the surface density of 
X-ray sources. In all cases, the normalization implied that there 
were $4\times10^{-7}$ X-ray sources with $F_{\rm X} \ge 3\times10^{-6}$ \phcms\
for every 1 \msun\ of stars.
}
\label{fig:dist}
\end{figure}

We then computed the surface density of sources as a function of offset
from \sgrastar\ and of the absolute values of the Galactic latitudes and 
longitudes, and plotted them in Figure~\ref{fig:dist}. To account for 
the variations in our sensitivity over the survey region, the contribution
of each source to the distribution was weighted by the inverse of the 
probability of detecting it (calculated as described in Appendix A).
We used the catalog of \citet{mun03a} to fill in the inner 8\arcmin\ 
around \sgrastar. We used the identical flux cuts in the \sgrastar\ field
as for the rest of the survey, which allowed us to include another
42 sources.

To compare the distribution of X-ray sources with that of the ordinary stellar 
population, we used models for the Galactic bulge and the central 
150 pc that were
derived from infrared maps by \citet{lzm02}, and the exponential model 
of the Galactic disk in Kent, Dame, \& Fazio (1991)\nocite{kdf91}. These
mass models are accurate to about 50\%. We 
modeled the central 150 pc with two components. We assumed that the 
central 15 pc is dominated by a spherical cluster with a mass density profile
\begin{equation}
\rho = {{\rho_c}\over{1 + (r/r_c)^n}}.
\end{equation}
For $r$$<$6 pc, we use $\rho_c$$=$$3.3\times10^{6}$ \msun\ pc$^{-3}$,
$r_c$=0.22 pc, and $n$=2. For $6<r<200$ pc,  $n$=3, $r_c$ 
remains the same, and $\rho_c$ is adjusted so that the function is 
continuous at $r=6$ pc. The total mass of the central cluster is 
$6\times10^7$~\msun. 

The rest of the central 150 pc is dominated by a disk-like distribution 
with a mass density
\begin{equation}
\rho = \rho_d r^{-n} \exp(-|z|/z_d).
\end{equation}
For $r$$<$120 pc, we take $\rho_d$=300 \msun\ pc$^{-3}$, $n$=$0.1$,
and $z_d$=45 pc. For 120$<$$r$$<$220 pc, we take $n$=3.5, leave
$z_d$ the same, and adjust $\rho_d$ so the function is continuous
at $r$=120 pc. For 220$<$$r$$<$2000 pc, we take $n$=10, and treat
the other parameters the same as above.
The total mass of this nuclear stellar disk is $1.4\times10^9$~\msun.

We model the Galactic bulge as a tri-axial ellipsoid of the form
\begin{eqnarray}
\rho = & \rho_{\rm bulge} e^{-r_s} \\
r_s = &\left[ (r_\bot)^{c_\|} + \left(\frac{|x|}{a_x}\right)^{c_\|} \right]^{1/c_\|} \\
r_\bot = & \left[ \left(\frac{|x|}{a_x}\right)^{c_\bot} + \left(\frac{|y|}{a_y}\right)^{c_\bot} \right]^{1/c_\bot}.
\end{eqnarray}
The axis defining $x,y,z$ is rotated 15\degree\ in east and 1\degree\ 
north from our line-of-sight.
The parameters are $a_x$=1100 pc, $a_y$=360 pc, $a_z$=220 pc, 
$c_\bot$=1.6, and 
$c_\|$=3.2, and $\rho_{\rm bulge}$=8 \msun pc$^{-3}$.
The total mass of the bulge is taken to be $10^{10}$ \msun.

Finally, we model the Galactic disk as a simple exponential,
\begin{equation}
\rho = \rho_0 \exp(-r/r_d) \exp(-|z|/z_d),
\end{equation}
where $r_d = 2.7$ kpc, $z_d = 200$ pc, and $\rho_0 = 5$ \msun\ pc$^{-3}$,
so that the total mass of the disk is $10^{11}$ \msun.

\begin{deluxetable*}{lcccccc}[thb]
\tablecolumns{7}
\tablewidth{0pc}
\tablecaption{Parameters of the $\log N - \log S$ Distribution\label{tab:ns}}
\tablehead{
\colhead{Field} & \colhead{$S_{\rm lim}$} & \colhead{Num.} & \colhead{Area} 
& \colhead{$\alpha$} & \colhead{$N_0$} & \colhead{$P_{\rm KS}$} \\
\colhead{} & \colhead{$10^{-6}$ \phcms} & \colhead{Sources} & 
\colhead{(arcmin$^{2}$)} & \colhead{} & \colhead{(arcmin$^{-2}$)} &
\colhead{}
} 
\startdata
Sgr B2 & 0.7 & 48 & 101 & 1.7$\pm$0.2 & 0.04 & 0.99 \\
Radio Arches & 2 & 28 & 97 & 1.1$\pm$0.2 & 0.17 & 0.69 \\
Shallow Survey & 6 & 142 & 2833 & 1.5$\pm$0.1 & 0.11 & 0.37 \\ [5pt]
\sgrastar\ & 0.8 & 232 & 152 & 1.4$\pm$0.1 & 0.24 & 0.20
\enddata
\tablecomments{The normalization of the $\log N - \log S$ distribution,
$N_0$ is listed for a fiducial flux of $3\times10^{-6}$ \phcms, to
match the spatial distribution in Figure~4. $P_{\rm KS}$ represents
the probability under a Kolgoromov-Smirnov
test of seeing the observed difference between the observed 
and model distribution assuming that they are identical, 
so that very small values would indicate a poorer match.}
\end{deluxetable*}

\begin{figure}
\centerline{\epsfig{file=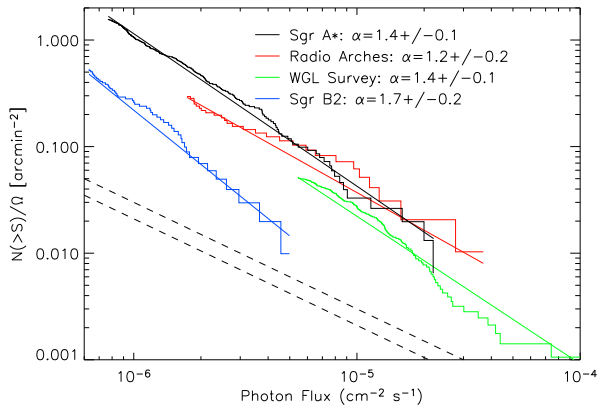,width=0.95\linewidth}}
\caption{
The number-flux distribution of sources in the Arches Region
(red lines), the Sgr B2 region (blue lines), and the shallow
survey (blue lines). The solid histograms have 
been corrected for the detection probability, and are only
plotted for sources brighter than the flux limits used in 
modeling the distributions. The solid lines are the best-fit
model distributions ($N(>S)\propto S^{-\alpha}$), with the 
slopes indicated in the upper-right of the figure. The dashed
lines represent the background AGN contributions taken from 
\citet{bra01} and \citet{man03}.
}
\label{fig:ns}
\end{figure}

We integrated the model stellar density along our line of sight toward
the Galactic center, using a lower limit of 6~kpc because we excluded
foreground sources from the profiles in Figure~\ref{fig:dist}, and 
an upper limit of 10~kpc because sources beyond this distance will be 
heavily absorbed and difficult to detect. 
We then compared these model surface densities to the observed surface
density of X-ray sources through a linear $\chi^2$ minimization.
We did not attempt to correct the surface density of X-ray 
sources to account for fact that the absorption column varies
as a function of longitude, latitude, and along our line of sight
in the image, because the uncertainty introduced by our failure to 
do so is smaller than the $\approx$50\% uncertainty in the mass model. 
The best-fit stellar surface densities are indicated by solid lines in
Figure~\ref{fig:dist}. The overall match is good, with 
$\chi^2/\nu < 1$. However, within 2\arcmin\ of \sgrastar\ (4.7 pc in 
projection) the number of X-ray sources is 2$\sigma$ larger than that 
expected from a simple scaling of the mass distribution inferred from 
the infrared (1.3$\pm$0.4 sources arcmin$^{-2}$ versus the predicted
0.5 sources arcmin$^{-2}$). This 
may be further evidence that X-ray sources are more concentrated near 
\sgrastar\ than ordinary stars are \citep[see also][]{mun05a}. 
The normalization of the fits imply that above a limit of 
$5\times10^{32}$~\ergs\ there are $(4 \pm 2)\times10^{-7}$ X-ray sources per 
solar mass of stars, where most of the uncertainty is in the mass 
models. We compare this to the expected density of X-ray
sources in \S4.

\subsection{Number-Flux Distribution\label{sec:ns}}

Spatial variations in the underlying population of X-ray sources could
be identified by examining the relative number of faint and bright 
X-ray sources. Therefore, we 
have computed the cumulative $\log{N}-\log{S}$ distributions in four regions:
the Arches field, the Sgr B2 field, and the general shallow survey 
(excluding the deep pointings), and the \sgrastar\ field 
\citep[see also][]{mun03a}. However, in this case, we required that 
each source had at least a 90\% chance of being detected, which was 
stricter than when we studied the spatial distribution.
We modeled the differential number-flux distribution using the method
described in Murdoch, Crawford, \& Jauncey (1973)\nocite{mcj73},
with slight modifications described in Appendix B to use Poisson 
statistics and knowledge of the average background rate.
As for the spatial distribution, we defined a flux
limit ($S_{\rm lim}$) for each region above which we could securely
detect sources over the largest possible area. For the 
$\log N - \log S$ distribution, we added the criterion that the flux 
limit allow us to measure the fluxes of sources at the 5$\sigma$ level.
Our choice of flux limits is described in more detail in Appendix B. 
The flux limits, area covered by each survey, and number of sources that were 
brighter than $S_{\rm lim}$ and located at points where sources with 
$S$=$S_{\rm lim}$ could be detected securely are 
listed in Table~\ref{tab:ns}. 

In Table~\ref{tab:ns}, we also list the best-fit slopes 
$\alpha$, the normalizations $N_0$ at $S_0$=$3\times10^{-6}$ \phcms\
(the fit had to be extrapolated for the shallow survey), 
and the probabilities $P_{\rm KS}$ that the model and observed 
distributions match according to the KS-test.
The increase in the normalization of the distributions is consistent
with the radial distribution of point sources in Figure~\ref{fig:dist}.
We find marginal, 1.8$\sigma$ evidence that the $\log{N}-\log{S}$ 
distribution is flatter in the Radio Arches region than in the Sgr B2
field or the survey as a whole, which would imply that the former
contains a larger proportion of high-luminosity sources.
Unfortunately, 
our constraints on the $\log{N}-\log{S}$ distribution are a bit poor, 
because the number of sources that meet the criterion of having 
$>$5$\sigma$ flux measurements (roughly, $>$ 25 photons) 
is small (see Fig.~\ref{fig:count}).
The best-fit distributions are plotted with solid lines in 
Figure~\ref{fig:ns}.

In Figure~\ref{fig:ns}, we also plot the expected distributions of 
background active galactic nuclei
(AGN). To convert between 
the photon fluxes we use here and the 2--8 keV energy flux 
that is reported in papers on the \chandra\ and \xmm\ deep fields, we 
assume that AGN are observed through an absorption column of 
$12\times10^{22}$ cm$^{-2}$ and that their spectra are described by a 
$\Gamma$=1.5 power law (the final results are not sensitive to a 
choice of $\Gamma$ between 1.2 and 1.8). We find that a source with a 
0.5--8.0 keV photon flux of $S_0 = 3\times10^{-6}$ \phcms\ in our 
survey would have a 2--8 keV flux of $6\times10^{-14}$ \ergcms\ if 
it were located at high Galactic latitude.  
Based on the number-flux distribution in the deep fields, 
expected density of background AGN is given by 
$N(>S) = N_0 (S/S_0)^{-1}$, where at $S_0 = 3\times10^{-6}$ \phcms, 
$N_0$=0.007 source arcmin$^{-2}$ in \citet{bra01}, and 
$N_0$=0.01 source arcmin$^{-2}$ in \citet{man03}.
This normalization is $<$10\% of that of the shallow survey in 
Table~\ref{tab:ns} (see also Fig.~\ref{fig:dist}). We conclude that 
$<$130 of the absorbed sources in our sample should be extra-galactic.

\section{Discussion}

We have presented a catalog of X-ray sources with fluxes between a
few $10^{31}$ and $10^{35}$ \ergs. The majority of the region
was covered with tiled pairs of 12~ks observations, in which 
the 50\% completeness 
limit is $1\times10^{33}$~\ergs. Two deeper exposures were more sensitive. 
A 50~ks toward the Radio Arches was complete to $4\times10^{32}$~\ergs, and
a 100~ks observations toward Sgr B2 was complete to $1\times10^{32}$~\ergs.
The sensitivity of the Radio Arches observation was a factor of two poorer
than would be naively expected based on the exposure time, because 
the Galactic center produces strong diffuse emission 
against which point sources are more difficult to detect.

Our survey encompasses a non-trivial fraction of the mass of stars in the
Galaxy. Integrating the \citep{lzm02} models for the stellar distribution 
over our survey area, we find that it encloses a stellar mass of 
$\sim$$10^{9}$ \msun, or 1\% of the Galactic value.
In Table~\ref{tab:pop}, we list the various classes of object that
could be detected as X-ray sources in our image if they were 
located near the Galactic center. We also include predictions for the
total number of each class of sources encompassed by the survey 
region, based on the references in the table. 
The estimates are based on the assumption that the star formation rate
is $\sim$1\% of the Galactic value, or 0.01 \msun\ yr$^{-1}$ \citep{fig04}. 
However, 
if the recent rate of star formation in the central 150 pc of 
the Galaxy is $\sim$10\% of the total Galactic rate, as is suggested
by indirect measurements of the Lyman-alpha flux in the region
\citep{cl89,fig99}, then WR/O stars, 
luminous pulsars, and high-mass X-ray binaries each could be an 
order-of-magnitude more numerous. Below we describe
the considerations that went into developing that table,
and some implications that the observed population of sources
have for understanding the evolution of the accreting binaries that 
make up the majority of our sample.

\subsection{Comparison to the Local Galactic X-ray Population}

We start by comparing the amount of X-ray flux per unit stellar mass
from the point sources with $L_{\rm X}$$>$$5\times10^{32}$~\ergs\ 
in our survey to that in the local Galaxy as identified by \citet{saz05}.
For a cumulative number-flux distribution
of the form $N(>L_{\rm X}) = N_0 (L_{\rm X}/L_{\rm X,0})^{-\alpha}$,
the specific luminosity produced by sources with 
$L_{\rm X,min} < L_{\rm X} < L_{\rm X,max}$ is
\begin{equation}
{\cal L}_{\rm X,tot} = \frac{\alpha N_0 L_{\rm X,0}}{\alpha -1} \left[ 
\left(\frac{L_{\rm X,min}}{L_{\rm X,0}}\right)^{-\alpha + 1} - 
\left(\frac{L_{\rm X,max}}{L_{\rm X,0}}\right)^{-\alpha + 1}\right],
\end{equation}
where $N_0$ is the normalization at a luminosity of $L_{\rm X,0}$ 
in units of sources per solar mass.
For the Galactic center, we have found that $\alpha = 1.5$ and that 
$N_0 = (4\pm2)\times10^{-7}$ sources \msun$^{-1}$ at 
$L_{\rm X,0} = 5\times10^{32}$ \ergs\
(Tab.~\ref{tab:ns}, Fig.~\ref{fig:dist}). So, the 
the specific luminosity of X-ray point sources  
with luminosities between $L_{\rm X,min} = 5\times10^{32}$ \ergs\ and
$L_{\rm X, max} = 10^{34}$ \ergs\ in our survey is 
${\cal L}_{\rm X,tot} = (5\pm2)\times10^{26}$ \ergs\ \msun$^{-1}$.
In the local Galaxy, \citet{saz05} find that    
$\alpha\approx 1.2$ and $N_0 = (6\pm2)\times10^{-4}$ sources \msun$^{-1}$ 
at $L_{\rm X,0} = 2\times10^{30}$ \ergs (where for $K$ in their Eq. 5, 
$N_0 = K/\alpha$, and we have estimated the uncertainty based on that 
of total specific luminosity for sources with 
$L_{\rm X} < 10^{34}$ \ergs). So, the specific luminosity of sources with
luminosities between $L_{\rm X,min} = 5\times10^{32}$ \ergs\ and
$L_{\rm X, max} = 10^{34}$ \ergs\ is locally
${\cal L}_{\rm X,tot} = (1.0\pm0.3)\times10^{27}$ \ergs\ \msun$^{-1}$. 
Therefore, the specific luminosities of X-ray sources with 
$5\times10^{32} < L_{\rm X} < 10^{34}$ \ergs\ 
in the local Galaxy and in the Galactic center are consistent within
their uncertainties.

To understand the population of X-ray sources at 
the Galactic center, it is notable that in the local Galactic neighborhood
all the X-ray sources with $10^{32} < L_{\rm X} < 10^{34}$ \ergs\ (2--10 keV) 
are cataclysmic variables (CVs) with magnetic white dwarfs and orbital 
periods of several hours (intermediate polars; Sazonov \etal\ 2005). 
Therefore, it is conceivable that most of the X-ray sources with
$L_{\rm X} < 10^{34}$~\ergs\ in our survey are magnetic CVs
\citep[see also][]{mun04b,lay05}. We would expect
CVs to have the same spatial distribution as the old ($\ga$ Gyr) 
population of stars, which dominate the infrared light from the
Galaxy, and indeed the distribution of X-ray sources is identical
to the inferred distribution of stars (Fig.~\ref{fig:dist}). As described
in \citet{mun04b}, intermediate polars have particularly hard, 
intrinsicly-absorbed spectra that are consistent with those of the
sources with $L_{\rm X} \la 10^{33}$ \ergs\ in Figure~\ref{fig:hit}.

Few CVs have $L_{\rm X} > 10^{33}$, and only one CV has 
been observed at $L_{\rm X} \ga 5\times10^{33}$~\ergs\ 
\citep[GK Per in outburst; e.g.,][]{krw79}, so we expect 
brighter sources to be more luminous objects such as LMXBs, HMXBs, WR/O stars 
in colliding-wind binaries, or pulsars (Table~\ref{tab:pop}).
The excess of bright X-ray sources observed in the
Radio Arches field (Fig.~\ref{fig:ns}), 
in which the Arches and Quintuplet clusters are
striking evidence recent active star formation
\citep{fig99}, can be explained if some of the bright sources
are HMXBs, WR/O stars, or young pulsars. Such sources have short 
lifetimes, and so 
should be concentrated near regions of active star formation. Such sources
also have softer spectra than intermediate polars in the 2--8 keV band,
and so could explain why the sources brighter than $\sim$$10^{33}$ \ergs\ 
in Figure~\ref{fig:hit} are systematically softer than the faint ones.

\begin{deluxetable}{lcccc}
\tablecolumns{5}
\tablewidth{0pc}
\tablecaption{X-Ray Sources in the Galactic Center\label{tab:pop}}
\tablehead{
\colhead{Object} & \colhead{$\log(L_{\rm X})$} & \colhead{Number} & 
\colhead{Number} & \colhead{References} \\
\colhead{} & \colhead{$\log({\rm erg~s}^{-1})$} & \colhead{in GC} & 
\colhead{Detectable} & \colhead{}
}
\startdata
CVs & 29.5--33.5 & $10^{5}$ & $10^{3}$ & [1,2] \\ 
WR/O Stars & 31--34 & $10^{3}$ & 10 & [3,4,5] \\
Pulsars & 29.3--35 & $10^{6}$ & 10 & [6,7] \\
LMXBs & 30--39 & $10^{3}$ & 10 & [8,9,10] \\
HMXBs & 31--38 & $10^{3}$ & 50 & [11]
\enddata
\tablerefs{[1] \citet{ver97}; [2] \citet{saz05}; [3] \citep{fig04}; 
[4] \citet{pol87}; [5] \citet{ber97}; [6] \citet{ba02}; [7] \citet{cl97}; 
[8] \citet{wij02}; [9] \citet{ll05}; 
[10] \citet{kon02}; [11] \citet{prp02}.}
\tablecomments{
We list order-of-magnitude estimates of the total population of various 
X-ray sources in our field, along with the number that should be detected 
in our survey.}
\end{deluxetable}

\subsection{Comparison to Theoretical Models}

Several binary population synthesis calculations have been carried 
out in order to interpret the population of X-ray sources in the
Galactic center described by \citet{wgl02} and \citet{mun03a}.
These models outline how the numbers of each class of X-ray sources
constrain various combinations of parameters in the binary evolution models
and assumptions about the physics of systems accreting at low rates.

For instance, \citet{prp02} suggested that several hundred of the X-ray
sources in the \citet{wgl02} survey could be neutron stars 
accreting from the winds of $>$3\msun\ binary companions 
\citep[see also][]{bt04}. The total 
numbers of wind-accreting neutron stars, and the fractions of systems 
with companions more and less massive than 
8\msun, varies by a factor of a few depending upon the the magnitudes 
of the kicks imparted to the neutron stars at birth. 
These theoretical predictions have motivated searches for infrared 
counterparts to the X-ray sources in the Galactic center surveys
\citep{lay05,ban05}. However,
\citet{ll05} suggested that accretion could be inhibited by the 
magnetospheres of the neutron stars \citep[see also][]{dp81}
at the low mass-transfer rates considered
by \citet{prp02}, so that wind-accreting neutron stars may not be 
luminous enough to be detected in our \chandra\ survey.

There also is theoretical disagreement as to whether CVs contribute 
significantly to the population of X-ray sources in our survey. 
\citet{ll05}, who base their calculations on the binary 
evolution code of Hurley, Tout, \& Pols (2002)\nocite{htp02}, suggest that CVs
are not luminous enough to be detected in large numbers from the Galactic 
center. However, this disagrees with similar calculations by \citet{rbh05},
who use the StarTrack code \citep{bel05}, and predict that significant numbers
of luminous CVs should be detectable from the Galactic center.
The main difference between the two codes is in their prescriptions for 
calculating the rate of mass transfer, which leads \citet{ll05} to predict 
mass transfer rates $\sim$100 times lower than those
used by \citet{rbh05} for identical systems with orbital periods of 
several hours (A. Ruiter, private communication). Systems
with orbits of several hours have the highest accretion rates
\citep[][see also Howell, Nelson, \& Rappaport 2001]{pat84}\nocite{hnr01}
and therefore are the most luminous in X-rays, which makes them the most
important contributors among CVs to our survey.
Our comparison with the local Galactic population of 
X-ray sources suggests that CVs are indeed both 
numerous and luminous enough to explain the population of X-ray sources 
in our image \citep[see also, e.g.,][]{ver97,ei99,srr05}, so our results 
could be taken as further, indirect evidence in support of the 
prescription used by \citet{bel05} and \citet{rbh05}.

Finally, \citet{bt04} and \citet{ll05} predict that there are a few
thousands LMXBs consisting of a neutron star accreting
from a white dwarf in our survey region. In these models, most of the 
neutron stars in these 
systems form through the accretion-induced collapse of an ONe white dwarf.
Under the assumptions of \citet{ll05}, a few percent of the LMXBs are 
persistently bright enough to be detected in our survey.
Moreover, most of these LMXBs should be transient, so if one assumes a 
standard duty cycle of $\sim$1\%, there should be $>$50 LMXBs in outburst
with $L_{\rm X} > 10^{36}$ \ergs\ in the field at any given time 
(and thousands in the Galaxy).  In contrast, there are only two 
persistent LMXBs this luminous in the field, 1E 1743.1--2843 and
1E 1740.7--2942 \citep{wgl02}, and several decades of occasional
X-ray observations have only revealed about a dozen transients with 
$L_{\rm X} \ga 10^{36}$ \ergs\ \citep[and only $\sim$150 in the rest of
the Galaxy; see, e.g.,][]{liu01,wij05}. The production of a large number of  
these transient LMXBs appears to be a common 
feature of models in which 
neutron stars can form through accretion-induced collapse 
(e.g., \S5.6 of Iben, Tutukov, \& Yungelson 1995)\nocite{ity95}. 
Under the models of \citet{bt04} 
and \citet{ll05}, possible ways to accommodate the small number of 
bright transients in our field are to assume that the efficiency with 
which the envelope of a star can be ejected during the common envelope 
phase is low so that many of the binaries merge rather than forming LMXBs,
or to assume that the accretion-induced collapse of an ONe white 
dwarf does not form a neutron star.

\section{The Future}

Further progress in understanding the natures of the X-ray sources 
near the Galactic center, and the consequential constraints on the
parameters input into binary evolution and population synthesis 
models, will be acquired through multi-wavelength observations of 
the region. X-ray observations will add to the population 
of transient LMXBs in the field \citep[e.g.,]{wgl02,sak05,mun05a,wij05}. 
Comparing infrared and X-ray surveys will reveal individual examples of 
WR/O binaries and HMXBs \citep[e.g.,]{ban05,mun05b}. The first pulsars 
near the Galactic center may be found by comparing 
radio and X-ray surveys \citep[see also][]{cl97}.

Finally, we note that this catalog will be improved upon greatly
in the next year, because a series of deep, 40~ks exposures of roughly
half of the survey area have been approved for the 2006 
\chandra\ observing cycle. We expect to increase the number of 
X-ray sources detected by a factor of $\approx$5, and improve the
uncertainties on the positions of the sources in the shallow 
survey from $\approx$1\arcsec\ to $<$0\farcs5. In the meantime, 
the current catalog provides the best available sample for 
studying the spatial and number-flux distributions of X-ray sources near the 
Galactic center, and for identifying their counterparts at 
other wavelengths.

\acknowledgments
We thank M. Morris for comments on the manuscript, K. Belczynski for
discussions about population synthesis models, and especially A. J. 
Ruiter and X. Liu for performing calculations to help clarify the differences
between various population synthesis codes. MPM and FEB acknowledge
support from NASA/SAO under grant AR5-6018A, and WQD acknowledges 
support from NASA/SAO under grant GO4-5010X. MPM also was supported 
under a Hubble Fellowship, and FEB under a Chandra Fellowship.
\appendix

\section{Completeness of the Survey}

In order to study the spatial and luminosity distributions of the 
X-ray sources, we need to calculate the limiting flux at which we
can confidently detect sources as a function of position in the 
survey. We performed synthetic star tests following the basic method 
of \citet{bau04}, with modifications to account for differences in 
our treatment of the photometry and the more complex layout of our survey. 

In order to produce $\sim$$10^{5}$ synthetic stars, we simulated 1000 
exposures of ObsID 2287 (12 ks exposure), and 100 each 
for ObsIDs 945 (50 ks exposure), and 944 (100 ks exposure). For each 
observation, we
removed events from within a circle circumscribing 92\% of the energy of
the PSF around each detected source. We then created images from the 
resulting event lists, and filled the ``holes'' in the image with a 
number of counts drawn from a Poisson distribution with a mean equal to
that of a surrounding annulus. We drew fluxes for synthetic stars from
a power-law distribution $N(>S) \propto S^{-\alpha}$ with a slope 
$\alpha = 1.5$, normalizations and flux limits chosen to match the
numbers and intensities of sources in each region (e.g., Tab.~\ref{tab:ns}).
We converted these fluxes to count rates using an exposure map 
generated for photons with $E = 3$~keV, and then drew net numbers of 
counts from Poisson distributions with those mean count rates. 
Each point source was assigned to a random position 
in the image. Next, we
obtained a model image of the PSF created by \program{mkpsf} when 
computing the photometry for
the the nearest detected source in the image, and used the PSF as the 
probability distribution to simulate the 2-dimensional image of 
the counts. This was placed into the composite image. Finally, we search 
the synthetic image for point sources 
using \program{wavdetect}, and tabulated which sources we placed into 
the image were detected by our search algorithm.

\begin{figure}
\centerline{\epsfig{file=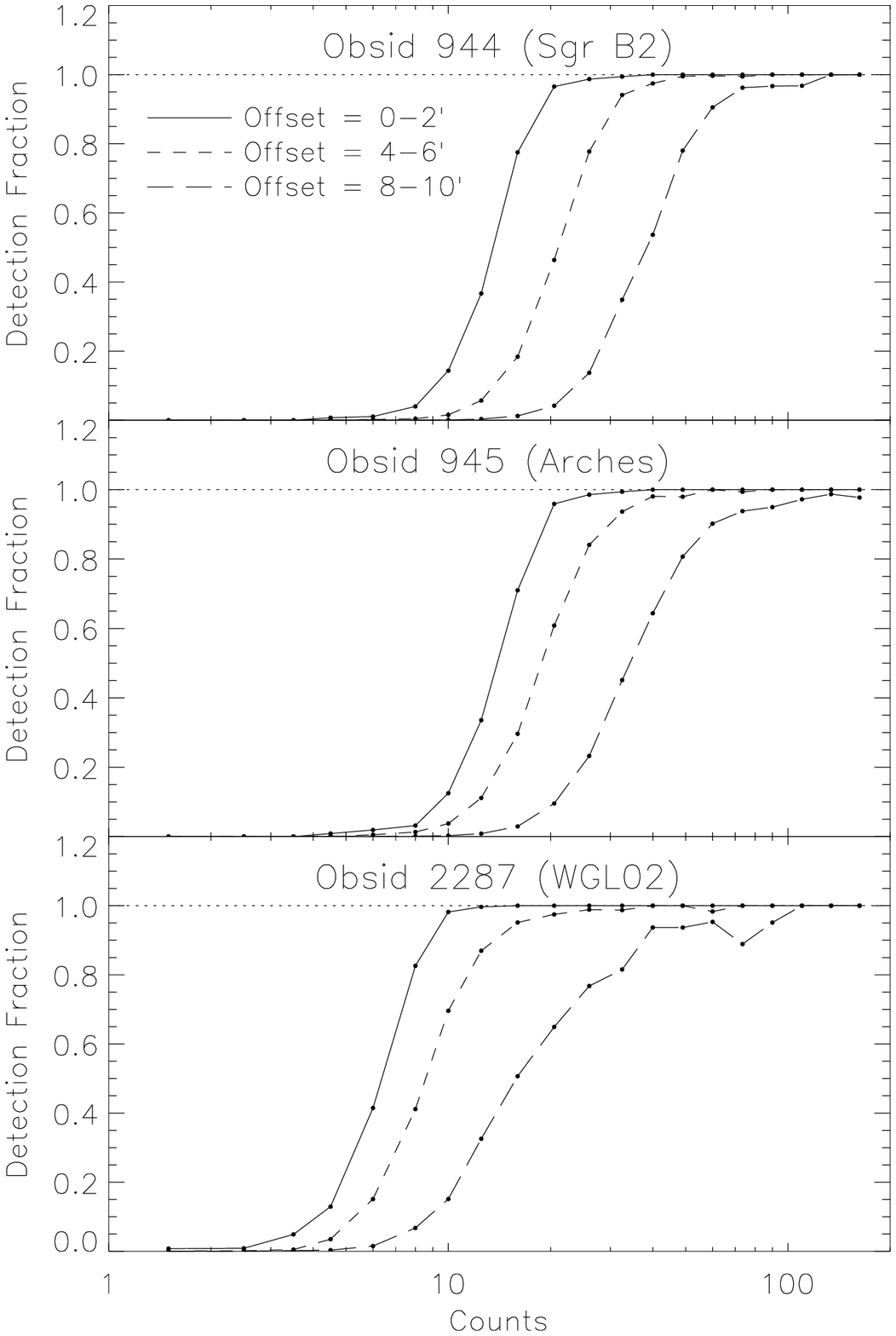,width=0.5\linewidth}}
\caption{
The probability of detecting a source as a function of count
rate (y-axis) and offset (line styles). 
The detection probability depends upon the survey region because of variations
in the diffuse X-ray background; the longer observations are 
background-limited. The probability at a given count rate decreases as
the offset increases, because the size of the PSF increases. We note
that the curve for the 8--10\arcmin\ offset in the shallow (WGL02) survey
was not used much, because sources in that survey almost always were found
closer to the aimpoint of at least one of the tiled pointings.}
\label{fig:fluxcurve}
\end{figure}

\begin{figure}
\centerline{\epsfig{file=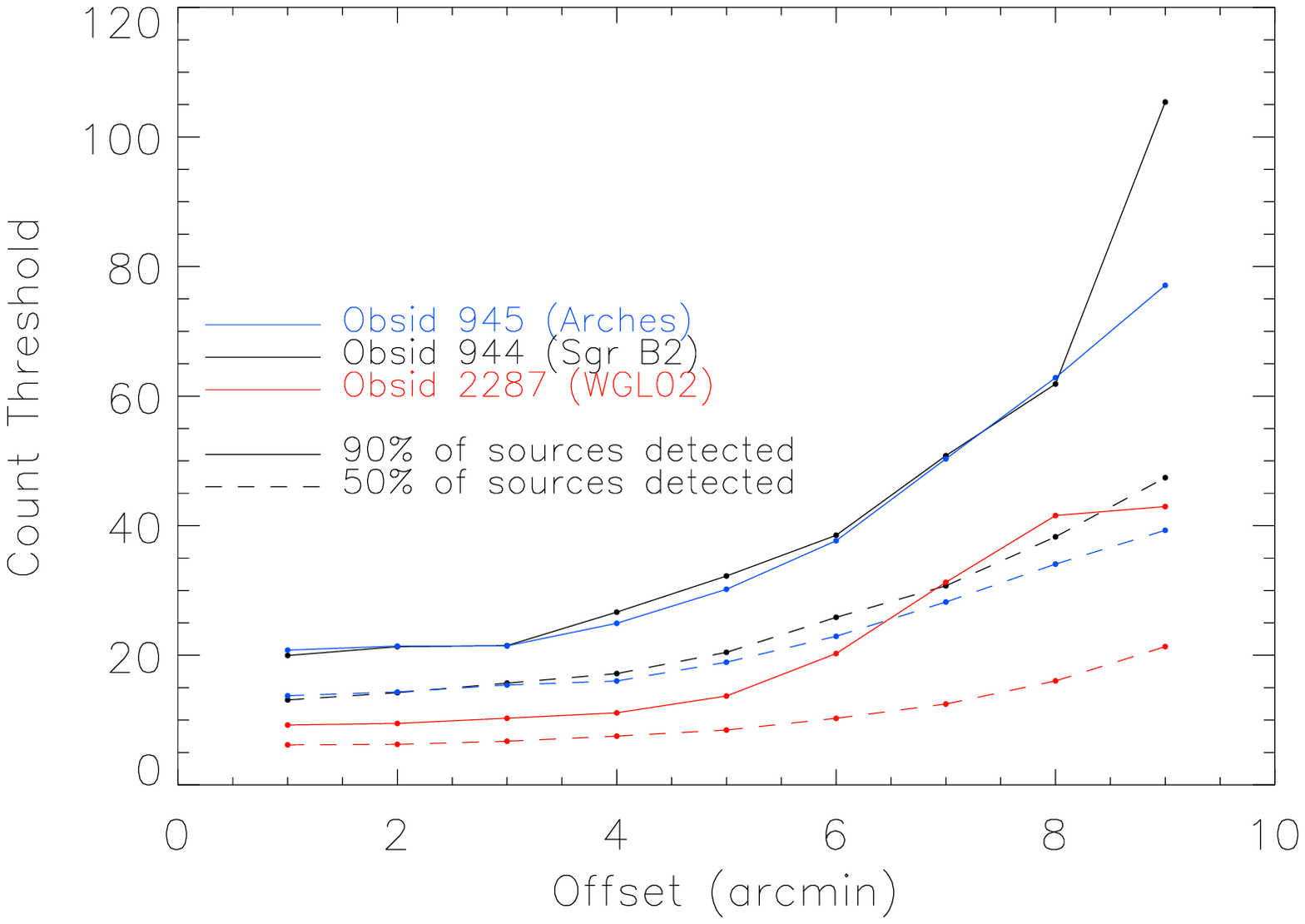,width=0.95\linewidth}}
\caption{
The number of counts required to detect a source in 50\% (dashed line)
and 90\% (solid line) of trials, as a function of offset from the
center of the cluster. In order of decreasing number of counts required,
the red line is for the Arches region, the 
blue line for Sgr B2, and the green line for the shallow survey. 
}
\label{fig:countthresh}
\end{figure}

We display the fraction of sources that are detected in 12, 50 and 100 ks
as a function of input flux and offset from the aim point in 
Figure~\ref{fig:fluxcurve}. The probability of detecting a source 
obviously increases with larger flux, because there is more 
signal-to-noise. For a given flux, the probability of detecting a source
is also larger at small offsets from the aim point, because farther from
the aim point counts are spread over a larger area by the PSF. This latter
trend can also be seen in Figure~\ref{fig:countthresh}, where we 
have plotted the number of counts above which a source will be detected
50\% and 90\% of the time as a function of offset from the aim point. 
As expected, longer observations enable us to reliably detect sources 
with lower fluxes. However, the total background counts from diffuse X-ray
emission toward the Galactic center also increases for longer exposures, so 
more counts are required to reliably detect a source. As a result, a 
factor of 5 increase in exposure from 10 to 50~ks yields only a factor of 
3 increase in sensitivity, which is by no means linear, but it is still
a faster improvement than the $t^{-1/2}$ trend that would be expected if 
we were background limited. 

\begin{figure}
\centerline{\epsfig{file=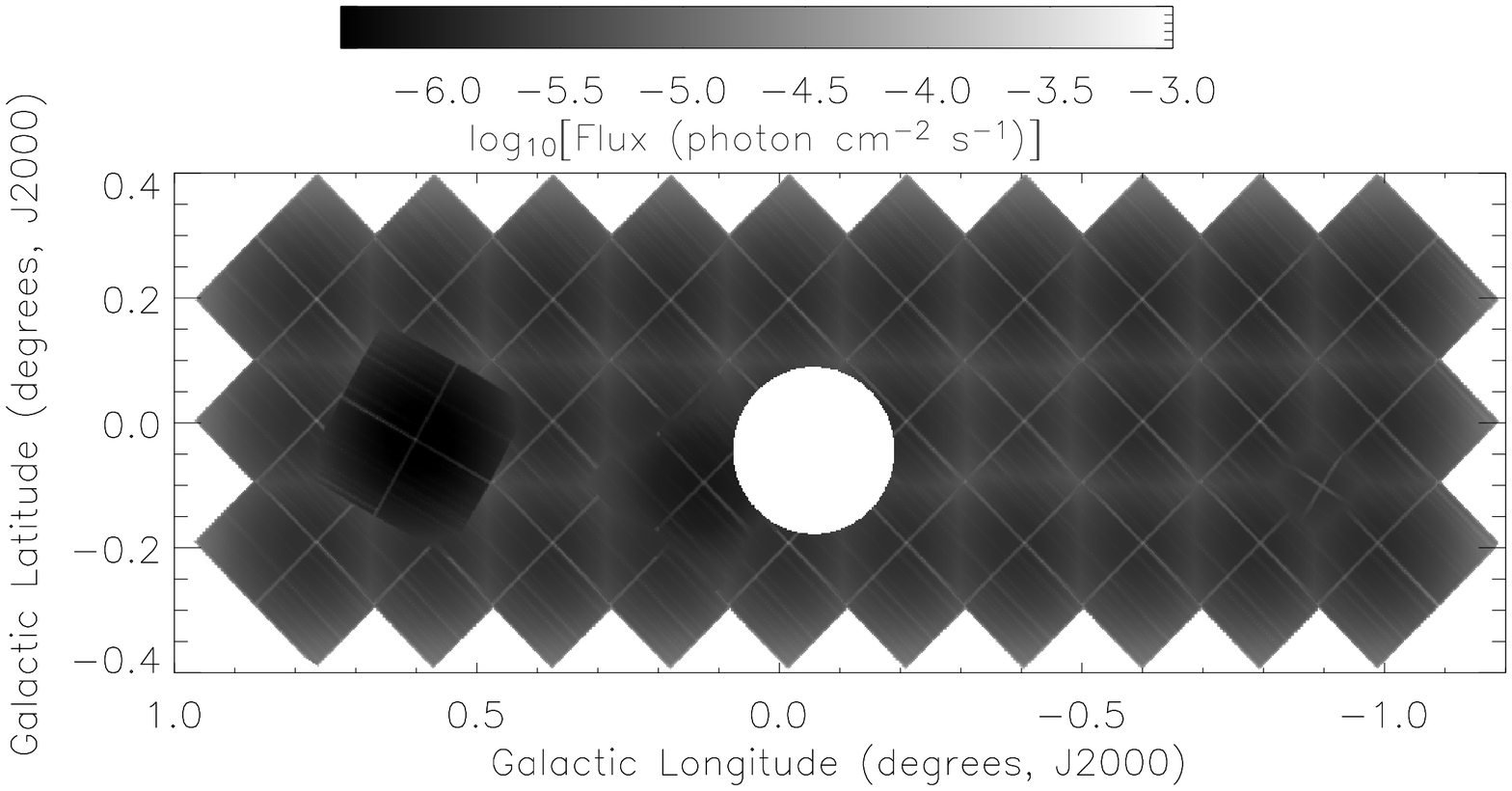,width=0.9\linewidth}}
\caption{
Map of the 50\% detection threshold over our survey. The large
white circle at the center represents the region covered by the 
deep survey in \citet{mun03a}. We do not include sources in that
region in our catalog.
}
\label{fig:threshmap}
\end{figure}

\begin{figure}
\centerline{\epsfig{file=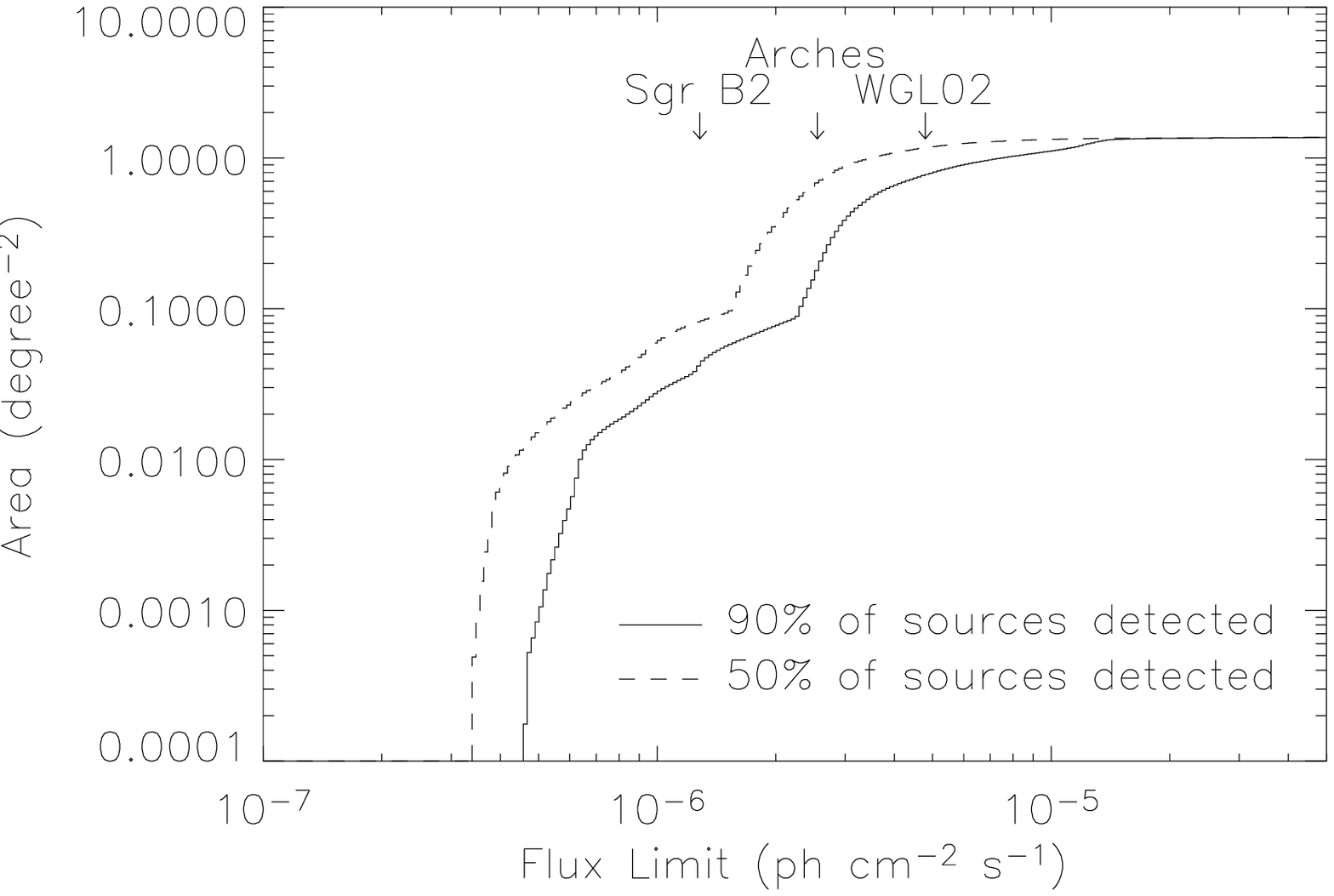,width=0.95\linewidth}}
\caption{
The area over which we were sensitive to sources of given fluxes. The
mean sensitivity of the Sgr B2, Arches Region, and shallow surveys are
indicated with arrows.
}
\label{fig:thresharea}
\end{figure}

We used the above simulations to make a map of the sensitivity of our 
survey. For each observation, we generated a sensitivity map by 
(1) computing the offset of each pixel from the aim point, (2) computing
the count threshold from Figure~\ref{fig:countthresh}, and (3) dividing
the count threshold by the exposure map (in units of cm$^{2}$ s) to 
obtain a flux. Then, to create a composite map for the full survey, for
each pixel we recorded the lowest value of the flux threshold from all of
the maps with exposure at that pixel. We note
that using this method, localized enhancements in the diffuse emission
that decrease our sensitivity have been averaged over offset from 
the aim point. Properly accounting for all of the variations in the diffuse
emission would require that we carry out our Monte Carlo simulations for
each exposure, which would be time consuming and would not change our 
results significantly. The resulting map is displayed
in Figure~\ref{fig:threshmap}. 
We also have tabulated the area over which our
survey is sensitive as a function of limiting flux, and displayed that 
in Figure~\ref{fig:thresharea}.

We have also used the simulations to compute the probability of detecting
each source. For each source, we determined the offset from the aim point
of the most sensitive exposure, and determined the count threshold
at that offset from Figure~\ref{fig:countthresh}. We then divided that 
count threshold by the exposure map from the most sensitive observation
at the location of the source to obtain a flux threshold. Again, this 
technique ignores local background variations that introduce systematic
errors in our computed thresholds for individual sources, but these 
errors should average out when considering large numbers of sources.
The median probability of detecting a Galactic center source was 57\%.
This is because the photons from faint sources are often lost to the 
wings of the PSF, and because 
the number-flux distribution is quite steep 
(see \S\ref{sec:ns}), so many faint sources are only detectable when  
Poisson fluctuations result in larger observed counts 
(this produces the Eddington bias when computing number-flux distribution).

The sensitivity maps and detection probabilities were used in computing
the spatial distribution in \S\ref{sec:dist} and flux distributions in 
\S\ref{sec:ns}. We also repeated the above process for the combined image 
created from 625~ks of exposure on the 17\arcmin$\times$17\arcmin\ field 
around \sgrastar, so that we could compare the results from the catalog
in \citet{mun03a} to the current one. We find that a source can be 
detected confidently with the fewest counts in the shallow survey 
(Fig.~\ref{fig:countthresh}), 
because the longer observations were background limited. Of course, the
observations were still more sensitive when the detection threshold was
considered as a function of flux (Fig.~\ref{fig:thresharea}).

\section{Modeling the Number-Flux Distribution}

We modeled 
the un-binned number-flux distributions using the technique described in 
\citet{mcj73}. We assumed that
the cumulative number-flux distribution could be described as a power law
$N(>S) \propto S^{-\alpha}$, where $S$ is the net number of counts for a 
source,
and that the number of observed counts could be described by a Poisson
distribution with mean rate $S+B$, where $B$ is the average number of 
background counts in the source extraction region (e.g., $B$=4.3 for the 
shallow survey).
The likelihood of observing sources with a distribution of 
total counts $C_i$ (which are all integers) is then: 
\begin{equation}
\sum_i \ln P(C_i) = \sum_i P_{{\rm det},i}
K^{-1} \int_0^\infty {{e^{-(S+B)} (S+B)^{C_i}}\over{C_i!}} N_0 S^{-(\alpha+1)} dS,
\label{eq:lognlogs}
\end{equation}
where $K$ is normalization over the range of count rates under consideration
($C_{\rm l}$, $C_{\rm u}$),
\begin{equation}
K = \int_0^\infty \sum_{C_{\rm l}}^{C_{\rm u}} P_{{\rm det},i} {{e^{-(S+B)} (S+B)^{C_i}}\over{C_i!}} N_0 S^{-(\alpha+1)} dS.
\end{equation}
The normalization of the power law
$N_0$ drops out of Equation~\ref{eq:lognlogs}, and so it was derived 
by setting the 
normalization $K$ for the best-fit $\alpha$ equal to the observed number
of sources. Finally, we compared the observed cumulative flux distribution 
to the
model distribution $P_i$ using a KS-test, to establish whether our power-law
model is consistent with the data.

Caution needs to be used in exercising these equations, as is described
in detail in \citet{mcj73}. First, the 
integrals over $S$ will diverge as $S$ approaches 0 unless a the count 
rate from a source is inconsistent with the background rate $B$ at 
the $\approx$5$\sigma$ level \citep{mcj73,wan04}. Therefore, we 
restricted our analysis to sources brighter than 
the 5$\sigma$ detection threshold for the average background level in 
each region. The mean background and count thresholds are listed in 
Table~\ref{tab:ns}.
In doing so, we were able to ignore the negligible contributions to the 
above integrals from beyond $S \approx (C_i) \pm 12(C_i)^{1/2}$.
Second, our source-detection algorithm was designed primarily to reject false
positives, and we find that false negatives occur for $\approx$20\%
of sources with count rates at the 5$\sigma$ level above background.
We use the factor $P_{{\rm det},i}$ in the above equations to account for 
the probability of detecting a source with a count rate $C_i$, which we 
determined from our Monte Carlo simulations. Third, in order to avoid
being biased by bright sources detected in regions with poor sensitivity, 
we only considered regions of the image in which there was 
a $>$90\% chance of identifying a source with a flux equal to the
5$\sigma$ detection threshold. 
We list in Table~\ref{tab:ns} the number of sources that met these 
criteria and the area over which we were sensitive.
Finally, we note that we
used average values for the background and the detection probability, 
even though both varied significantly over the regions covered by 
each set of observations. The total counts $C_i$ used
above were the sum of the net counts derived from our photometry and
the average background, which was then rounded to the nearest integer.
This was necessary to ensure that the 
integrand in Equations B1 and B2 were monotonic functions of $C_i$ 
\citep[see also][]{wan04}. 

Our approach to modeling the $\log{N}-\log{S}$ distribution
takes into account the possible Eddington bias, although it ignores
the vast majority of faint sources. Different approaches are
possible, and should yield similar results. For instance, \citet{bau04}
used Monte Carlo simulations to estimate the corrections to the flux 
required to offset both the Eddington bias 
and biases introduced by their method for deriving the photometry for 
each source, 
assumed that the photometric uncertainties were negligible, and 
analytically computed $\alpha$ from a maximum-likelihood distribution using  
corrected fluxes and the equations in \citet{mcj73}.
We have not implemented this technique, because our procedure for computing
the photometry for each source was computationally
prohibitive to incorporate it into our Monte Carlo simulations. We also
note that their resulting slope could be somewhat biased, because they have
assumed a $\log{N}-\log{S}$ distribution in computing the flux correction,
thereby pre-determining the effect of the Eddington bias. 

\citet{wan04}
has presented a method that is almost equivalent to ours, in which he 
applied a 
redistribution matrix to convert a model distribution into an observed 
distribution, and then found best-fit parameters for the model using the 
chi-squared and Cash minimization techniques implemented in \program{XSPEC} 
\citep{arn96}. If the model and un-binned fluxes are compared using the Cash 
statistic \citep{cash79}, the techniques are equivalent, although the
use of a response matrix converts the integral in Equation~\ref{eq:lognlogs}
into a sum. Using binned data and 
a chi-squared test obviously requires enough sources per bin that their
numbers are approximately distributed as a Gaussian. That technique has 
the advantage of using tools that X-ray astronomers are familiar with,
although it is conceptually more complicated than our method.

\clearpage
\begin{landscape}

\begin{deluxetable*}{lcccccccccccc}[htb]
\tabletypesize{\scriptsize}
\tablecolumns{13}
\tablewidth{0pc}
\tablecaption{Catalog of Point Sources within 2\degree$\times$0.8\degree\ of the Galactic Center\label{tab:cat}}
\tablehead{
\colhead{Source} & \colhead{ra} & \colhead{dec} & \colhead{Unc.} & 
\colhead{Obsid} &
\colhead{Offset} & \colhead{$T_{\rm exp}$} & \colhead{$C_{\rm net}$} & 
\colhead{$P_{\rm det}$} & \colhead{HR0} & 
\colhead{HR2} & \colhead{$F_{\rm X}$ $10^{-7}$} & \colhead{Flags} \\
\colhead{(CXO J)} & \multicolumn{2}{c}{(J2000)} & \colhead{(arcsec)} & 
\colhead{} & \colhead{(arcmin)} & \colhead{(ks)} & 
\colhead{} & \colhead{} & \colhead{} & \colhead{} & \colhead{(\phcms)} & 
\colhead{} 
}
\startdata
$174204.8-295003$ & $265.52017$ & $-29.83436$ & 1.5 &     2289 & 11.3 &  11.6 & $  15.4^{+ 8.0}_{- 6.9}$ & 0.00 & $-0.52_{-0.42}^{+0.39}$ & $-9.00$ &   60.2 &      sf \\
$174206.8-293634$ & $265.52844$ & $-29.60947$ & 1.0 &     2290 &  6.4 &  11.6 & $   4.2^{+ 4.1}_{- 2.7}$ & 0.25 & $-9.00$ & $ 1.00_{-0.76}$ &   22.8 & \nodata \\
$174208.5-294436$ & $265.53574$ & $-29.74361$ & 1.0 &     2290 &  6.5 &  11.6 & $   8.2^{+ 5.5}_{- 4.0}$ & 0.27 & $-1.00^{+0.50}$ & $-9.00$ &   24.8 &       f \\
$174210.4-293639$ & $265.54358$ & $-29.61107$ & 1.0 &     2290 &  5.7 &  11.6 & $  14.4^{+ 6.8}_{- 5.7}$ & 0.95 & $-9.00$ & $ 0.29_{-0.45}^{+0.46}$ &   64.7 & \nodata \\
$174216.0-293756$ & $265.56689$ & $-29.63249$ & 0.9 &     2290 &  3.9 &  11.6 & $   8.7^{+ 5.0}_{- 4.5}$ & 0.81 & $ 0.15_{-0.73}^{+0.70}$ & $-9.00$ &   25.9 & \nodata \\ [5pt]
$174216.1-293732$ & $265.56723$ & $-29.62580$ & 0.9 &     2290 &  4.2 &  11.6 & $  11.6^{+ 5.8}_{- 5.3}$ & 0.99 & $-9.00$ & $ 0.67_{-0.34}^{+0.33}$ &   57.0 & \nodata \\
$174217.8-293715$ & $265.57425$ & $-29.62091$ & 0.9 &     2290 &  4.1 &  11.6 & $  10.6^{+ 5.6}_{- 5.0}$ & 0.96 & $ 1.00_{-1.09}$ & $-0.35_{-0.55}^{+0.58}$ &   36.0 & \nodata \\
$174218.6-293931$ & $265.57761$ & $-29.65884$ & 0.9 &     2290 &  2.7 &  11.6 & $   6.7^{+ 4.3}_{- 4.0}$ & 0.94 & $ 1.00_{-1.09}$ & $ 0.20_{-0.90}^{+0.80}$ &   25.2 & \nodata \\
$174219.2-294333$ & $265.58029$ & $-29.72591$ & 0.9 &     2290 &  4.1 &  11.6 & $   4.5^{+ 3.8}_{- 3.0}$ & 0.17 & $-9.00$ & $-1.00^{+0.74}$ &   14.1 & \nodata \\
$174220.1-293526$ & $265.58389$ & $-29.59068$ & 0.9 &     2290 &  5.3 &  11.6 & $   4.5^{+ 3.9}_{- 3.0}$ & 0.06 & $ 1.00_{-0.89}$ & $-9.00$ &   13.5 &       s \\ [5pt]
$174220.1-293905$ & $265.58414$ & $-29.65145$ & 0.9 &     2289 &  2.6 &  11.6 & $  10.9^{+ 5.3}_{- 5.3}$ & 0.99 & $-1.00^{+0.55}$ & $ 1.00_{-0.87}$ &   36.9 &       f \\
$174221.3-294250$ & $265.58887$ & $-29.71409$ & 0.9 &     2290 &  3.3 &  11.6 & $   2.7^{+ 2.7}_{- 2.3}$ & 0.02 & $-9.00$ & $-1.00^{+1.08}$ &    7.7 & \nodata \\
$174221.3-294647$ & $265.58902$ & $-29.77988$ & 1.1 &     2286 &  6.9 &  11.6 & $   8.6^{+ 5.3}_{- 4.8}$ & 0.63 & $ 1.00_{-1.13}$ & $ 0.46_{-0.64}^{+0.55}$ &   40.4 & \nodata \\
$174222.8-294118$ & $265.59518$ & $-29.68858$ & 0.9 &     2290 &  2.0 &  11.6 & $   3.7^{+ 3.2}_{- 2.8}$ & 0.34 & $-9.00$ & $-0.00_{-0.98}^{+0.99}$ &   14.2 & \nodata \\
$174223.3-293950$ & $265.59735$ & $-29.66397$ & 0.9 &     2290 &  1.7 &  11.6 & $   5.8^{+ 3.9}_{- 3.7}$ & 0.69 & $-9.00$ & $-0.35_{-0.65}^{+0.77}$ &   20.6 & \nodata \\ [5pt]
$174224.2-293412$ & $265.60095$ & $-29.57001$ & 1.0 &     2290 &  6.2 &  23.2 & $  12.0^{+ 7.1}_{- 5.3}$ & 0.53 & $ 1.00_{-0.85}$ & $-0.02_{-0.56}^{+0.59}$ &   29.7 & \nodata \\
$174224.6-294333$ & $265.60275$ & $-29.72592$ & 0.9 &     2283 &  3.6 &  11.6 & $   8.6^{+ 5.1}_{- 4.4}$ & 0.73 & $-0.01_{-0.97}^{+0.99}$ & $-1.00^{+0.64}$ &   22.3 & \nodata \\
$174227.5-292602$ & $265.61487$ & $-29.43403$ & 1.2 &     2283 &  8.1 &  11.6 & $  66.5^{+13.9}_{-13.1}$ & 0.89 & $-0.55_{-0.18}^{+0.18}$ & $-0.20_{-0.65}^{+0.76}$ &  204.3 &      sf \\
$174228.3-294157$ & $265.61816$ & $-29.69925$ & 0.9 &     2290 &  1.8 &  23.2 & $   3.3^{+ 4.2}_{- 3.0}$ & 0.11 & $-9.00$ & $-0.12_{-0.88}^{+1.13}$ &   10.1 & \nodata \\
$174228.4-293431$ & $265.61850$ & $-29.57547$ & 1.0 &     2290 &  5.7 &  23.2 & $   9.8^{+ 5.7}_{- 5.4}$ & 0.09 & $-1.00^{+0.42}$ & $ 1.00_{-1.33}$ &   15.4 &       f \\ [5pt]
$174228.5-293736$ & $265.61883$ & $-29.62687$ & 0.9 &     2290 &  2.7 &  23.2 & $   7.8^{+ 5.8}_{- 5.7}$ & 0.75 & $-9.00$ & $ 1.00_{-1.13}$ &   20.3 & \nodata \\
$174230.0-293949$ & $265.62524$ & $-29.66380$ & 0.9 &     2290 &  0.4 &  23.2 & $   4.3^{+ 5.1}_{- 4.1}$ & 0.10 & $-9.00$ & $-9.00$ &    9.3 & \nodata \\
$174230.3-294713$ & $265.62646$ & $-29.78708$ & 1.1 &     2283 &  7.0 &  11.6 & $   8.6^{+ 5.4}_{- 4.7}$ & 0.17 & $ 1.00_{-0.92}$ & $-1.00^{+0.51}$ &   24.3 & \nodata \\
$174230.6-294320$ & $265.62769$ & $-29.72239$ & 0.9 &     2290 &  3.1 &  23.2 & $  20.2^{+ 9.2}_{- 7.7}$ & 0.99 & $ 0.07_{-0.54}^{+0.62}$ & $-1.00^{+0.98}$ &   46.9 & \nodata \\
$174231.4-294336$ & $265.63089$ & $-29.72681$ & 0.9 &     2290 &  3.4 &  23.2 & $< 8.7$ & 0.00 & \nodata & \nodata &    8.9 & \nodata
\enddata
\tablecomments{The full table will be available online, and is 
and is on
{\tt http://astro.ucla.edu/\~mmuno/sgra/shallow\_survey\_catalog.txt}.}
\end{deluxetable*}

\clearpage
\end{landscape}

\end{document}